\documentclass[11pt,a4paper]{amsart}
\usepackage[T1]{fontenc}
\usepackage[utf8]{inputenc}
\usepackage{lmodern,amsmath,amssymb,amsthm,booktabs}
\usepackage[margin=26mm]{geometry}
\usepackage{hyperref}
\hypersetup{colorlinks=true,linkcolor=blue,citecolor=blue,urlcolor=blue,
 pdftitle={Excess charge in the Engel-Dreizler model of atoms and molecules},
 pdfauthor={Rafael Benguria and Heinz Siedentop}}
\newtheorem{theorem}{Theorem}[section]
\newtheorem{lemma}[theorem]{Lemma}
\newtheorem{proposition}[theorem]{Proposition}
\newtheorem{corollary}[theorem]{Corollary}
\theoremstyle{remark}
\newtheorem{remark}[theorem]{Remark}
\newcommand{\R}{\mathbb R}
\newcommand{\dd}{\,\mathrm d}
\newcommand{\asinh}{\operatorname{arsinh}}
\newcommand{\atanh}{\operatorname{artanh}}
\newcommand{\E}{\mathcal E}
\newcommand{\D}{\mathcal D}

\newcommand{\norm}[1]{\lVert #1\rVert}
\title[Excess charge in the Engel--Dreizler model]{Excess charge in the Engel--Dreizler model of atoms and molecules}
\author{Rafael Benguria}
\address{Instituto de F\'isica\\
  Pontificia Universidad Cat\'olica de Chile\\
  Av.~Vicu\~na Mackenna 4860\\ Macul\\ 7820436, Santiago\\ Chile}
\email{rbenguri@uc.cl}
\author{Heinz Siedentop}
\address{Mathematisches Institut\\ Ludwig-Maximilians-Universit\"at
  M\"unchen\\ Theresienstra\ss e 39\\ 80333 M\"unchen\\ and Munich
  Center for Quantum Science and Technology (MCQST)\\
  Schellingstr. 4\\ 80799 M\"unchen\\ Germany}
\email{h.s@lmu.de}
\thanks{Support by the Deutsche Forschungsgemeinschaft (DFG), grant
  EXC-2111-390814868 and DFG TRR 352, grant 470903074, and FONDECYT
  (Chile) Project \# 124--1863 is gratefully acknowledged.
  RB thanks the Mathematical Institute of LMU for hospitality, and HS
  thanks the Institute of Physics of the Pontificia Universidad
  Cat\'olica de Chile for hospitality. The authors acknowledge the use
  of OpenAI's ChatGPT and Google's Gemini for mathematical discussions,
  the development and checking of arguments, and assistance in preparing
  the manuscript. The authors take full responsibility for the final
  text and proofs.}
\date{Draft, 6 November 2026}
\begin{document}
\begin{abstract}
We prove an explicit excess-charge bound for the exchange-free
Engel--Dreizler functional of atoms and molecules. If
$\alpha\lambda\le1/1233$ and $\alpha^3Z_k^2\le1/137$ for each nucleus,
every finite-mass electronic minimizer whose total energy, including
nuclear repulsion, is nonpositive satisfies
$N-\sum_kZ_k<(5913/1000)\lambda^{3/2}K$.
For the gradient coefficient $\lambda=1/9$, this gives
$N-\sum_kZ_k<0.219K$. The energy hypothesis holds automatically for atoms
and for local minima of the Born--Oppenheimer energy at which the electronic
infimum is attained. The proof uses no symmetry of the density, restriction
on the total nuclear charge, or minimum nuclear separation.
We combine molecular Sommerfeld estimates with a convex nonlinear
comparison in the variable $F(p)$. The failure of the latter comparison
at very high density is represented by a nonnegative Newton potential.
A quartic coercivity estimate, obtained from nuclear repulsion and a
signed localized Hardy inequality, controls the total mass of this source.
The chemical potential is retained throughout. Exact covariance yields
the invariant parameter domain, and all finite numerical inequalities
are reduced to rational arithmetic.
\end{abstract}
\maketitle

\section{Introduction}\label{sec:introduction}

The excess charge $Q:=N-Z$ of an atomic anion is the number of
bound electrons $N$ in excess of the nuclear charge $Z$. Experimentally,
no stable doubly or more highly charged isolated atomic anions are known;
see Massey \cite{Massey1976,Massey1979}. Explaining the observed bound
$Q\le1$ from a microscopic model of ions remains a challenge to
mathematical physics. A weaker formulation, the ionization conjecture,
asks for a bound on $Q$ independent of $Z$.

Simple density functionals already capture parts of this picture.
Thomas--Fermi theory does not bind negative ions, whereas its
Fermi--Amaldi modification permits at most one excess electron.
For the Thomas--Fermi--Weizs\"acker functional, Benguria and Lieb
\cite{BL} proved charge-independent bounds for atoms and bounds
proportional to the number of nuclei for molecules. Solovej established
such a uniform bound in reduced Hartree--Fock theory and subsequently
in Hartree--Fock theory with exchange \cite{Solovej1991,Solovej2003}.
These results give $Q=O(1)$ as $Z\to\infty$, with a universal bound
valid for all $Z$, but do not establish the sharp value $1$.
More recently, an OpenAI preprint \cite{OpenAI2026} established a
corresponding uniform bound for strictly bound states of the full
nonrelativistic many-electron Coulomb Hamiltonian with two spin states.
For $K$ fixed nuclei with real charges $Z_k\ge1$ at arbitrary distinct
positions, the result is $N\le Z_{\rm tot}+CK$, with a universal
constant $C$. In particular, it gives $Q=O(1)$ for atoms as
$Z\to\infty$, and hence the analogue of Solovej's Hartree--Fock
bound for the full $N$-particle model. Earlier quantitative bounds
for that model grew with $Z$. The new result does not provide the
sharp bound $Q\le1$.

For heavy atoms, relativistic effects must also be taken into account.
Although the highest velocities occur near the nucleus, their influence
extends to the outer shells and to chemical and optical properties;
the colour of gold is a familiar example \cite{Pyykko1988}.
Rigorous excess-charge results for relativistic microscopic models are
comparatively scarce. Schulz \cite[Section~2.2]{Schulz2025} reviews
this problem and proves, for the fermionic Chandrasekhar model,
\[
 Q\le\tfrac12(\sqrt2-1)Z+O(Z^{1/3}).
\]
Here the remainder is uniform when $\alpha Z$ stays a fixed distance
below the critical coupling $2/\pi$; more precisely, his
Theorem~2.2.1 assumes $a\alpha Z<2/\pi$ for a fixed $a>1$.

Motivated by these questions, we consider a relativistic statistical
model with a gradient correction. The relativistic Thomas--Fermi
model of Vallarta and Rosen \cite{VallartaRosen1932} suffers from
collapse at a point nucleus, as discussed by Jensen \cite{Jensen1933};
see also Gomb\'as \cite[\S~14, pp.~120~ff.]{Gombas1949}.
Engel and Dreizler later obtained a relativistic analogue of the
Thomas--Fermi--Weizs\"acker functional through a formal derivation
from the Hartree--Fock approximation to quantum electrodynamics
\cite{ED87}. They subsequently studied neutral and positively charged
atomic solutions numerically \cite{ED88}. We omit the exchange term
and ask whether this model admits an explicit excess-charge bound
uniform in the nuclear charges within a specified coupling domain.

The exterior estimates used below follow the Benguria--Lieb strategy,
with particular use of Solovej's refined Sommerfeld comparison after
Lemma~10 of \cite{Solovej}: see Lemma~11, Proposition~12, and Lemma~13
there. His Proposition~14 illustrates how the improved comparison
reduces the excess-charge constant. Uniform estimates for a broader
family of nonrelativistic Thomas--Fermi--Weizs\"acker functionals were
obtained by Benguria and Siedentop \cite{BS2025}.

There are already rigorous excess-charge estimates for the
Engel--Dreizler model.  Chen and Siedentop proved, for the functional
without exchange, a molecular bound
$N<2.56\sum_k Z_k$ \cite{CS}.  For the atomic functional including
exchange, Chen, Frank, and Siedentop obtained
$N\leq 2Z+O(Z^{2/3})$ as $Z\to\infty$ with
$\kappa=Z/(c\sqrt{\lambda})$ and $\lambda$ fixed, where $c$ is the speed of light in the Hartree units used there \cite{CFS}.
These results bound the total electron number by a quantity of order
the nuclear charge.  The present work estimates the difference
$Q=N-\sum_k Z_k$ by an explicit constant times the number of nuclei.

In the molecular argument, the repulsion of the nuclei is essential
to the energy estimate used to impose restrictions on individual
charges.  This use of nuclear repulsion has a precedent in the
ultrarelativistic Thomas--Fermi--Weizs\"acker stability analysis of
Benguria, Loss, and Siedentop \cite{BLS}.  Their homogeneous model
differs from the massive Engel--Dreizler functional considered here.
The relevant electrostatic localization inequalities originate in
the work of Lieb and Yau \cite{LY}; we use the formulation of Lieb,
Loss, and Siedentop \cite{LLS}.  Stability estimates for the
Engel--Dreizler functional were also developed by Siedentop
\cite{Siedentop}.

Our main result, Theorem~\ref{thm:main}, bounds the excess charge by
$(5913/1000)\lambda^{3/2}K$ in an explicitly stated coupling domain.
At $\lambda=1/9$ the constant per nucleus is $219/1000$.
The atomic result is a direct corollary of the molecular theorem.
The molecular energy hypothesis is the nonpositivity of the actual total
energy, including nuclear repulsion. A virial identity makes this hypothesis
automatic at local minima of the Born--Oppenheimer energy, whenever the
electronic infimum is attained.

There are two main steps in the proof. First, the exterior comparison
estimates depend only on $\alpha\lambda$ and on the function
$H(x)=\sum_{k=1}^K|x-R_k|^{-1}$, which is harmonic away from the
nuclear positions. They do not depend on individual nuclear charges.
Second, a nonlinear comparison between the density and the screened
potential is allowed to have a nonnegative Newton-potential correction.
Its source is supported where the local Fermi momentum exceeds an
explicit exponentially large cutoff. A quartic energy estimate makes the
electron number on this set small enough to absorb the correction.
The nuclear repulsion enters only in this second step, through a restriction
on each nuclear charge. Unequal charges are handled by an exact averaging
over subsets of nuclei.

Section~\ref{sec:model} states the result and specifies the rescaling from
Hartree units used throughout the paper.
Section~\ref{sec:variational} supplies the variational framework and
the exact covariance. The exterior estimates, coercivity bound, and
nonlinear comparison are proved in Sections~\ref{ext:section},
\ref{coer:section}, and~\ref{sec:nonlinear}, respectively.
Section~\ref{sec:charge} completes the charge estimate, and
Section~\ref{sec:nuclear} treats the nuclear energy condition and the
geometric factor $K$. The appendices give the algebra behind the localization
constant and a reproducible specification of the finite numerical checks.
The large-coupling regime at fixed $\alpha$ is outside the present theorem
and is left to a separate study.

\section{The functional and the main results}\label{sec:model}

In Hartree units, $\hbar=m_{\mathrm e}=e^2/(4\pi\varepsilon_0)=1$ and
$c=\alpha^{-1}$. We rescale the spatial and density variables by
\[
 x=cx_{\rm H},\qquad R_k=cR_{k,{\rm H}},\qquad
 p(x)=c^{-1}p_{\rm H}(x/c),\qquad
 \rho(x)=c^{-3}\rho_{\rm H}(x/c),
\]
where the subscript $\mathrm H$ denotes Hartree variables. This change
preserves the particle number. The Hartree energy and the functional
$\E$ defined in \eqref{eq:energy} are related by
\[
 \E_{\rm H}[p_{\rm H};\mathbf R_{\rm H},\mathbf Z]
 =c^2\E[p;\mathbf R,\mathbf Z].
\]
We omit the positive overall factor $c^2$: it rescales the energy but
preserves the corresponding minimizers, their particle numbers, and
the sign of the total energy. Thus the formulas below use
$\hbar=m_{\mathrm e}=c=1$, with Coulomb coupling $\alpha$.
Let $\alpha,\lambda>0$, and set
\begin{align}
 f(p)^2&=\frac p{\sqrt{1+p^2}}
       +\frac{2p^2}{1+p^2}\asinh p,
 &F(p)&=\int_0^pf(s)\dd s,\label{eq:f}\\
 T(p)&=\sqrt{1+p^2}-1,
 &\tau(p)&=8\int_0^p s^2T(s)\dd s.\label{eq:T}
\end{align}
The coefficient $2$ in $f^2$ is the standard Engel--Dreizler coefficient.
Write
\[
 P=F(p),\qquad \rho=\frac{p^3}{3\pi^2},\qquad
 N=\int_{\R^3}\rho,\qquad Z_{\rm tot}=\sum_{k=1}^K Z_k,
 \qquad Q=N-Z_{\rm tot}.
\]
The nuclei have pairwise distinct positions $R_1,\ldots,R_K$ and charges
$Z_k>0$. Define
\[
 V(x)=\sum_{k=1}^K\frac{Z_k}{|x-R_k|},\qquad
 U_{\rm nn}=\alpha\sum_{1\le k<l\le K}
                  \frac{Z_kZ_l}{|R_k-R_l|},
\]
so that each pair is counted once. The total energy is
\begin{equation}\label{eq:energy}
 \E[p;\mathbf R,\mathbf Z]
 =\frac{3\lambda}{8\pi^2}\int|\nabla F(p)|^2
 +\frac1{8\pi^2}\int\tau(p)
 -\alpha\int V\rho+\alpha\D[\rho]+U_{\rm nn},
\end{equation}
where
\[
 \D[\rho]=\frac12\iint
       \frac{\rho(x)\rho(y)}{|x-y|}\dd x\dd y.
\]
Apart from the gradient term and the fixed nuclear repulsion,
\eqref{eq:energy} is the relativistic Thomas--Fermi functional introduced
by Vallarta and Rosen \cite{VallartaRosen1932}. In contrast to the
nonrelativistic Thomas--Fermi kinetic energy, the local term involving
$\tau$ does not stabilize the energy against collapse at a point nucleus:
without a gradient correction, the electronic energy is unbounded below.
Various ad hoc modifications were proposed to address this difficulty,
including the addition of the nonrelativistic Weizs\"acker term; see
Gomb\'as \cite[\S~14, pp.~120~ff.]{Gombas1949} for the early approaches.
The advantage of the first term in \eqref{eq:energy} is that its form
comes from Engel and Dreizler's formal QED-based gradient expansion
\cite{ED87}, rather than an ad hoc choice. The coefficient $\lambda$
is retained as a parameter; its standard value is $1/9$.

There is no exchange term. The finite-mass energy class is
$p\ge0$, $p\in L^3\cap L^4$, $F(p)\in D^{1,2}$, and $\D[p^3]<\infty$.

Here $D^{1,2}(\R^3)$ is the completion of $C_c^\infty(\R^3)$ in the
gradient norm, and all unmarked integrals are over $\R^3$.
The first two terms of \eqref{eq:energy} are denoted by $T_{\rm W}[p]$
and $T_{\rm TF}[p]$, respectively.

An \emph{electronic minimizer} always means an attained infimum for fixed
$\mathbf R$ and $\mathbf Z$ over this energy class, with either
$N\le M$ for a prescribed finite $M$, $N=M$ for a prescribed finite $M$,
or no constraint on $N$. In the last case we explicitly require finite
mass. We use $N$ for the actual electron number of the minimizer, including
when the constraint is relaxed. The argument applies to each such minimizer
and does not require uniqueness. The underlying existence and ionization
theory for the exchange-free functional is developed in~\cite{CS}.

\begin{theorem}[Excess charge]\label{thm:main}
Let $K\ge1$, let $R_1,\ldots,R_K\in\R^3$ be pairwise distinct, and let
$Z_1,\ldots,Z_K>0$. Suppose
\begin{equation}\label{eq:invariantdomain}
 \alpha\lambda\le\frac1{1233},\qquad
 \alpha^3Z_k^2\le\frac1{137}\quad(1\le k\le K).
\end{equation}
If $p$ is a finite-mass electronic minimizer and its actual total energy
satisfies
\begin{equation}\label{eq:sublevel}
 \E[p;\mathbf R,\mathbf Z]\le0,
\end{equation}
then
\begin{equation}\label{eq:main}
 \boxed{\displaystyle N-Z_{\rm tot}<\frac{5913}{1000}\lambda^{3/2}K.}
\end{equation}
The bound is independent of the nuclear positions and of the total nuclear
charge. No lower bound on the nuclear separations is assumed.
\end{theorem}

\begin{corollary}[The physical gradient coefficient]\label{cor:physical}
For $\lambda=1/9$, $0<\alpha\le1/137$, and $\alpha Z_k\le1$ for every
nucleus, an electronic minimizer satisfying \eqref{eq:sublevel} obeys
\begin{equation}\label{eq:physicalbound}
 \boxed{\displaystyle N-Z_{\rm tot}<\frac{219}{1000}K.}
\end{equation}
\end{corollary}
\begin{proof}
The stated assumptions imply \eqref{eq:invariantdomain}, since
$\alpha^3Z_k^2=\alpha(\alpha Z_k)^2\le1/137$.
Moreover $(5913/1000)(1/9)^{3/2}=219/1000$.
\end{proof}

\begin{corollary}[Atoms]\label{atom:corollary}
Let $Z,\alpha,\lambda>0$ satisfy
\[
 \alpha\lambda\le\frac1{1233},\qquad
 \alpha^3Z^2\le\frac1{137}.
\]
Every finite-mass atomic minimizer, with free mass, a relaxed mass
constraint, or an attained exact mass constraint, satisfies
\[
 N-Z<\frac{5913}{1000}\lambda^{3/2}.
\]
In particular, for $\lambda=1/9$, $0<\alpha\le1/137$, and
$\alpha Z\le1$, one has $N-Z<219/1000$.
\end{corollary}

\begin{proof}
For one nucleus the nuclear repulsion is zero. In the free and
relaxed problems, the zero density is admissible, so the minimizing
energy is nonpositive. For an exact particle number $M>0$, choose
a smooth nonnegative compactly supported momentum $q$ with
$\int q^3/(3\pi^2)=M$, and set $q_L(x)=L^{-1}q(x/L)$, with
the nucleus placed at the origin. The particle number is unchanged.
As $L\to\infty$, both kinetic energies are $O(L^{-2})$ and both
Coulomb contributions are $O(L^{-1})$. Thus $\mathcal E[q_L]\to0$,
and an attained exact-mass minimum also has nonpositive energy.
The case $M=0$ is immediate. The result now follows from
Theorem~\ref{thm:main} with $K=1$ in its invariant parameter domain.
The stated physical assumptions imply that domain because
$\alpha^3Z^2=\alpha(\alpha Z)^2\le1/137$ and
$\alpha\lambda\le1/1233$.
\end{proof}

The nuclear repulsion does not change the electronic Euler equation. It
supplies the energy control needed in the molecular proof. In particular,
Theorem~\ref{thm:main} applies at a local minimum of the Born--Oppenheimer
energy whenever the electronic infimum is attained; this follows from
Corollary~\ref{cor:BO} below. The theorem does not assert existence of such
a nuclear minimum. For an arbitrary externally fixed configuration,
\eqref{eq:sublevel} remains an explicit hypothesis.

The numerical coefficient is an upper bound established by the present
comparison argument; no optimality is claimed. Uniformity in the charges
means uniformity within \eqref{eq:invariantdomain}. This domain places no
restriction on $Z_{\rm tot}$ when the number of nuclei is allowed to vary.

\section{Variational framework and covariance}\label{sec:variational}

\subsection{The Euler equation and regularity}

We record the prerequisites to distinguish qualitative regularity from a
uniform maximum-density estimate. No uniform bound on $\norm p_\infty$
will be used.

\begin{lemma}\label{lem:domain}
The finite-mass energy class is equivalent to $P=F(p)\in H^1(\R^3)$,
$P\ge0$. In this class $p\in L^6$ and $\rho\in L^1\cap L^2$.
\end{lemma}
\begin{proof}
The elementary bound $f(p)^2\le3p$ implies $F(p)^2\le(4/3)p^3$.
Conversely, $F(p)\asymp p^{3/2}$ for $0\le p\le1$, and
$F(p)\ge c p$ for $p\ge1$. Thus
\[
 p^3\le C(P^2+P^3),\quad p^4\le C(P^{8/3}+P^4),\quad
 p^6\le C(P^4+P^6).
\]
Sobolev embedding gives these integrability statements. Interpolation and
the Hardy--Littlewood--Sobolev inequality give finite direct energy; each
nuclear attraction is finite by $\rho\in L^1\cap L^2$.
\end{proof}

\begin{lemma}\label{lem:EL}
Every nonzero finite-mass electronic minimizer as defined in Section~\ref{sec:model} satisfies
\begin{equation}\label{eq:EL}
 \frac{3\lambda}{4}\Delta P=j(p)(T(p)-t),\qquad
 j(p)=\frac{p^2}{f(p)},\qquad
 t=\alpha\varphi-\eta,\qquad
 \varphi=V-\rho*|\cdot|^{-1},
\end{equation}
with $\eta\ge0$. For a relaxed cutoff $M$, $\eta(N-M)=0$;
for a free minimizer $\eta=0$. Also
\begin{equation}\label{eq:Poisson}
 \Delta t=4\pi\alpha\rho-4\pi\alpha\sum_kZ_k\delta_{R_k}
\end{equation}
in distributions.
\end{lemma}
\begin{proof}
Extend the energy evenly to real $P\in H^1$, with
$\mathsf n(P)=F^{-1}(|P|)^3/(3\pi^2)$ and
$\mathsf k(P)=\tau(F^{-1}(|P|))/(8\pi^2)$.
Their derivatives at $P>0$ are
$\mathsf n'=p^2/(\pi^2f)$ and $\mathsf k'=p^2T/(\pi^2f)$.
They extend continuously through zero, with growth at most
$C(|P|+|P|^2)$ and $C(|P|+|P|^3)$, respectively. The scalar bounds below
give $|\mathsf n''(P)|\le C(1+|P|)$.
Sobolev embedding, Hardy's inequality at each of finitely many nuclei,
and the Coulomb inequality make the mass and energy $C^1$ on $H^1$.
At a nonzero minimizer the mass derivative is nonzero. The multiplier rule
for a relaxed constraint gives $\eta\ge0$ and \eqref{eq:EL}.

The exact and relaxed mass infima agree. To see this, multiply a relaxed
competitor $P$ by smooth compactly supported cutoffs taking values in
$[0,1]$ and tending to one, with convergence to $P$ in $H^1$.
The resulting particle numbers do not exceed the original one, and
the continuity established above gives convergence of the masses and
energies, including the nuclear and Hartree terms. For each such
approximation, restore the missing mass with a smooth nonnegative
momentum cloud
\[
 q_L(x)=L^{-1}q\bigl((x-a_L)/L\bigr),
\]
where $q\in C_c^\infty(\R^3)$ is nonnegative and normalized to that
missing mass. Choose $|a_L|/L\to\infty$
so that the cloud and the compactly supported approximation have
disjoint supports and their distance tends to infinity.
The cloud's kinetic energies are $O(L^{-2})$, its direct energy is
$O(L^{-1})$, and its interactions with the nuclei and the original
density tend to zero. First letting $L\to\infty$ and then removing the
cutoff proves equality of the two infima. Nuclear repulsion is fixed
in this argument.
An attained exact-mass minimizer is consequently a relaxed minimizer as
well. Free variations give $\eta=0$ in the free case.
\end{proof}

\begin{lemma}\label{lem:regularity}
Every nonzero finite-mass minimizer has $P,p$ continuous and strictly
positive on $\R^3$, smooth off the nuclei, with
\[
 P\in W^{2,q}_{\rm loc}\ (1<q<3),\qquad
 \norm p_\infty<\infty,\qquad p(x)\longrightarrow0
 \quad\text{uniformly as }|x|\to\infty.
\]
The Hartree potential belongs to $C_0\cap L^\infty$.
\end{lemma}
\begin{proof}
The last assertion follows first from $\rho\in L^1\cap L^2$, by splitting
the Coulomb kernel and approximating the density in both norms.
The quotient $p^2/(f(p)F(p))$ extends to $3/2$ at zero and is at most
$4+2p$. The Euler equation therefore gives
\[
 -\Delta P\le C(1+p)V P.
\]
The coefficient lies locally in $L^{8/5}$, since $p\in L^4$ and each
$|x-R_k|^{-1}$ is locally in $L^{8/3}$, with $1/4+3/8=5/8$.
We use the local $L^2$-to-$L^\infty$ estimate for nonnegative weak
subsolutions of $-\Delta u\le a u$: if $a\in L^q(B_2(x))$ with
$q>3/2$, then
\[
 \sup_{B_1(x)}u\le C\|u\|_{L^2(B_2(x))},
\]
where $C$ depends only on $q$ and $\|a\|_{L^q(B_2(x))}$.
It applies here with $q=8/5$. On unit balls far from all nuclei
the coefficient norms are uniformly bounded. Consequently,
\[
 \sup_{B_1(x)}P\le C\norm P_{L^2(B_2(x))}\longrightarrow0
\]
gives boundedness and decay. The right side of \eqref{eq:EL} now lies
locally in $L^q$ for $q<3$, proving continuity and the asserted regularity.
Writing $\Delta P=a(x)P$ gives $a\in L^q_{\rm loc}$ for some $q>3/2$;
Harnack's principle and bootstrapping complete the proof.
All constants in this qualitative argument may depend on the configuration.
\end{proof}

\subsection{Exact covariance and the parameter domain}\label{scov:section}

The following simultaneous change of lengths, charges, and parameters
preserves the form of the functional. Subscripts on $\E$ display its
parameters explicitly.

\begin{proposition}[Exact spatial covariance]
\label{scov:covariance}
Let $s>0$, put $x=sy$, and define
\begin{gather}
 \widetilde p(y)=p(sy),\qquad
 \widetilde R_k=R_k/s,\qquad \widetilde Z_k=Z_k/s^3,\label{scov:change}\\
 \widetilde\lambda=\lambda/s^2,\qquad
 \widetilde\alpha=\alpha s^2.
\end{gather}
Then
\begin{equation}\label{scov:energy}
 \boxed{\displaystyle
 \mathcal E_{\lambda,\alpha,\mathbf Z,\mathbf R}[p]
 =s^3\mathcal E_{\widetilde\lambda,\widetilde\alpha,
                 \widetilde{\mathbf Z},\widetilde{\mathbf R}}
                 [\widetilde p].}
\end{equation}
Moreover,
\begin{equation}\label{scov:mass}
 N=s^3\widetilde N,\qquad Q=s^3\widetilde Q,
 \qquad \widetilde\eta=\eta,
\end{equation}
where the chemical potential is written as $-\eta$. In particular, the
sign of the total energy is preserved, as is the sign of $\eta$.
\end{proposition}

\begin{proof}
The gradient, local kinetic, and Hartree integrals acquire the respective
factors $s$, $s^3$, and $s^5$. Since
$\widetilde\rho(y)=\rho(sy)$ and
\[
 \sum_k\frac{Z_k}{|sy-R_k|}
 =s^2\sum_k\frac{\widetilde Z_k}{|y-\widetilde R_k|},
\]
the electron--nuclear integral also acquires the factor $s^5$ when written
with the transformed charges. The nuclear repulsion transforms consistently:
\[
 s^3\widetilde\alpha
 \frac{\widetilde Z_k\widetilde Z_l}
      {|\widetilde R_k-\widetilde R_l|}
 =\alpha\frac{Z_kZ_l}{|R_k-R_l|}.
\]
These identities prove \eqref{scov:energy}. The particle-number formula
follows by integration, and the charge formula follows because
$Z_{\rm tot}=s^3\widetilde Z_{\rm tot}$. Finally, energy and particle number
have the same prefactor $s^3$, so the Lagrange multiplier is unchanged.
Equivalently, for the screened potentials,
\[
 \widetilde\varphi(y)=s^{-2}\varphi(sy),\qquad
 \widetilde\alpha\widetilde\varphi(y)=\alpha\varphi(sy).
\]
\end{proof}

The transformation is a bijection between the corresponding admissible
classes. It therefore transports electronic minimizers with exact,
relaxed, or free particle number, with the particle-number constraint
rescaled as in \eqref{scov:mass}. It also transports joint minimization over
the electronic density and nuclear positions. Pairwise distinct nuclear
positions remain pairwise distinct.

Useful invariant quantities are
\begin{equation}\label{scov:invariants}
 \gamma=\alpha\lambda,\qquad
 \kappa_k=\frac{\alpha Z_k}{\sqrt\lambda},\qquad
 \chi_k=\alpha^3Z_k^2=\gamma\kappa_k^2,
\end{equation}
and
\[
 \frac{Z_k}{\lambda^{3/2}},\qquad
 \frac{R_k}{\sqrt\lambda},\qquad
 \frac{Q}{\lambda^{3/2}}.
\]
Taking $s=\sqrt\lambda$ sets $\widetilde\lambda=1$, but changes the
coupling, charges, and nuclear geometry simultaneously. The factor
$\lambda^{3/2}$ in an excess-charge bound is consistent with this exact
covariance.

\begin{lemma}[Reduction of the parameter domain]\label{lem:reduction}
To prove \eqref{eq:main} under \eqref{eq:invariantdomain}, it suffices to
prove it under
\begin{equation}\label{eq:normalizeddomain}
 0<\alpha\le\frac1{137},\qquad
 \gamma:=\alpha\lambda\le\gamma_0:=\frac1{1233},\qquad
 \alpha Z_k\le1\quad(1\le k\le K).
\end{equation}
The electronic minimizing and total-energy hypotheses are preserved.
\end{lemma}
\begin{proof}
Take $s=(137\alpha)^{-1/2}$ in Proposition~\ref{scov:covariance}. Then
\[
 \widetilde\alpha=\frac1{137},\qquad
 \widetilde\lambda=137\alpha\lambda\le\frac19,
 \qquad
 \widetilde\alpha\widetilde Z_k=\sqrt{137\alpha^3Z_k^2}\le1.
\]
The transformed parameters satisfy \eqref{eq:normalizeddomain}. Multiplying
$\widetilde Q<(5913/1000)\widetilde\lambda^{3/2}K$ by $s^3$ gives
\eqref{eq:main}. The remaining assertions follow from the covariance.
\end{proof}

In fact, \eqref{eq:invariantdomain} is precisely the set of parameters
that can be mapped into \eqref{eq:normalizeddomain} by this covariance:
the possible scaling factors satisfy
$\alpha\max_k Z_k\le s\le(137\alpha)^{-1/2}$.
In particular, there is no separate restriction on $\alpha$ in
Theorem~\ref{thm:main}.

\section{Universal exterior estimates}
\label{ext:section}

In this section $\gamma=\alpha\lambda\le\gamma_0=1/1233$.
We use the functions in \eqref{eq:f}--\eqref{eq:T}, the Euler equation
\eqref{eq:EL}, and the qualitative regularity of Lemma~\ref{lem:regularity}.
Away from the nuclei,
\begin{equation}\label{ext:system}
 \frac{3\lambda}{4}\Delta P=j(p)(T(p)-t),\qquad
 \Delta t=\frac{4\alpha}{3\pi}p^3.
\end{equation}
No restriction on the nuclear charges, no separation condition, and no
global comparison between $P$ and $t$ is used in this section.

The molecular geometry is encoded by the unweighted function
\begin{equation}\label{ext:H}
 H(x)=\sum_{k=1}^K\frac1{|x-R_k|}.
\end{equation}
It is harmonic on $\R^3\setminus\{R_1,\ldots,R_K\}$, where
\begin{equation}\label{ext:Hidentities}
 \Delta H=0,\qquad
 |\nabla H|\le\sum_k|x-R_k|^{-2}\le H^2,\qquad
 \tau_H:=\frac{|\nabla H|^2}{H^4}\in[0,1].
\end{equation}

\subsection{Scalar estimates and the form inequality}

\begin{lemma}\label{ext:scalar}
For $p>0$, $f$ is increasing, and
\begin{equation}\label{ext:scalarbounds}
 f(p)^2\le p+2p^3,\qquad
 \frac{f(p)^2}{p}\le\frac{17}{8},\qquad
 0<\frac{pf'(p)}{f(p)}\le\frac7{10}.
\end{equation}
In particular, $j(p):=p^2/f(p)$ is strictly increasing, and
$10/17\le F(p)/(pf(p))\le1$. Also
\begin{equation}\label{ext:Aconstant}
 \frac34\frac{f(p)^2}{p\bigl(2-pf'(p)/f(p)\bigr)}
 \le\frac{43}{50}.
\end{equation}
\end{lemma}
\begin{proof}
Set $v=p/\sqrt{1+p^2}$, $u=v^2$, and
$L=\atanh v=\asinh p$. If $h=f^2$, then
\[
 h=v+2v^2L,\qquad
 ph'=v(1+v^2)+4v^2(1-v^2)L>0,
\]
and
\[
 7h-5ph'=v(2-5u)+2v^2(10u-3)L.
\]
For $u\ge3/10$, use $L\ge v$ to bound the last expression below by
$v(2-11u+20u^2)>0$. For $u\le3/10$, use
$L\le v/(1-u)$, taking account of the negative coefficient, to obtain
$v(2-13u+25u^2)/(1-u)>0$. Both quadratics have negative discriminants.
Thus $pf'/f=ph'/(2h)\le7/10$. Integrating the logarithmic derivative
bound gives $f(s)\ge f(p)(s/p)^{7/10}$ for $s\le p$, and hence
$F(p)\ge(10/17)pf(p)$. Monotonicity gives $F(p)\le pf(p)$.
The inequality $\asinh p\le p$ gives $h\le p+2p^3$.
Putting $w=(1+p^2)^{-1/2}$ also gives
$h/p\le w+2-2w^2\le17/8$.

The localization estimate \eqref{ext:Aconstant} follows from the
algebraic argument in Appendix~\ref{app:localization}.
\end{proof}

\begin{lemma}\label{ext:formlemma}
With $A=(43/50)\lambda$, the quadratic-form inequalities
\begin{equation}\label{ext:form}
 -A\Delta+T(p)-t\ge0,\qquad
 -A\Delta+\tfrac12p^2-t\ge0
\end{equation}
hold. Their coefficient does not depend on the number or positions of the
nuclei.
\end{lemma}
\begin{proof}
Let $v(P)=f(p)/p^2$ and $D(P)=-v'(P)>0$. Testing
\eqref{ext:system} with a smooth function $b$ supported away from all nuclei
and completing a square gives
\begin{align*}
 \int(T-t)b^2
 &=\frac{3\lambda}{4}\int v\Delta P\,b^2\\
 &=\frac{3\lambda}{4}\int D
       \left|b\nabla P-\frac vD\nabla b\right|^2
   -\frac{3\lambda}{4}\int\frac{v^2}{D}|\nabla b|^2.
\end{align*}
Here
\[
 \frac{v^2}{D}=\frac{f(p)^2}{p(2-pf'/f)}.
\]
The first assertion follows from \eqref{ext:Aconstant}; the second uses
$T(p)\le p^2/2$. The inequalities extend by density. Removing finitely
many points by capacity cutoffs, with Hardy's inequality at each point,
does not change the coefficient. In the application below, all test balls
are already disjoint from the nuclei.
\end{proof}

\subsection{The molecular Sommerfeld barrier}

\begin{lemma}\label{ext:sommerfeldscalar}
Let $k,\ell>0$ and put
\begin{equation}\label{ext:thetaM}
 \theta=\frac{31}{40},\qquad M_{\rm S}=\frac{692881}{57600},\qquad
 b_\ell(r)=\frac{M_{\rm S}^2}{k^2r^4}
       \left[1-(\ell/r)^\theta\right]^{-4},\quad r>\ell.
\end{equation}
Then $\Delta b_\ell\le k b_\ell^{3/2}$ in three dimensions.
\end{lemma}
\begin{proof}
For arbitrary $M,\theta>0$, put $s=(\ell/r)^\theta$. Direct differentiation
gives
\begin{align*}
 \Delta b_\ell-kb_\ell^{3/2}
 ={}&\frac{M^2}{k^2}r^{-6}(1-s)^{-6}\\
 &\quad\times\{12-M+4(\theta^2+7\theta-6)s
                   +(16\theta^2-28\theta+12)s^2\}.
\end{align*}
For the values in \eqref{ext:thetaM}, the relevant quadratic satisfies
\begin{equation}\label{ext:sommerfeldsquare}
 12+\frac{41}{400}s-\frac9{100}s^2
 =\frac{692881}{57600}
       -\frac9{100}\left(s-\frac{41}{72}\right)^2.
\end{equation}
The asserted sign follows. The barrier diverges at $r=\ell$ and tends to
zero at infinity.
\end{proof}

\begin{proposition}\label{ext:potentialbarrier}
For every $0<\beta<1$ and every point away from the nuclei,
\begin{equation}\label{ext:U}
 t(x)\le U_\beta(H(x)),\qquad
 U_\beta(h):=a_\beta h^4+b_\beta h^2,
\end{equation}
where
\begin{equation}\label{ext:ab}
 a_\beta=\frac{9\pi^2M_{\rm S}^2}
                 {128\alpha^2(1-\beta^\theta)^4},\qquad
 b_\beta=\frac{43\lambda\pi^2}{50\beta^2}.
\end{equation}
In particular, the coefficients are exactly those of the one-centre
Sommerfeld comparison.
\end{proposition}
\begin{proof}
For a fixed smearing radius $\ell>0$, use the normalized Dirichlet ball
ground state
\[
 g_\ell(y)=\frac{\sin(\pi|y|/\ell)}{\sqrt{2\pi\ell}|y|}
               \mathbf1_{\{|y|<\ell\}},\qquad
 \|g_\ell\|_2=1,\qquad
 \|\nabla g_\ell\|_2^2=\frac{\pi^2}{\ell^2}.
\]
Set $\widetilde\rho=\rho*g_\ell^2$ and
$v=t*g_\ell^2-A\pi^2/\ell^2$. Testing \eqref{ext:form} with translates
of $g_\ell$ and applying concavity of $\rho^{2/3}$ gives
\[
 v\le\frac12(3\pi^2)^{2/3}\widetilde\rho^{2/3}.
\]
On
$D_\ell:=\{x:|x-R_k|>\ell\text{ for all }k\}$,
\[
 \Delta v=4\pi\alpha\widetilde\rho
       \ge \mathsf{k}(v_+)^{3/2},\qquad \mathsf{k}=\frac{2^{7/2}\alpha}{3\pi}.
\]
Let $r_k=|x-R_k|$ and $B_\ell(x)=\sum_k b_\ell(r_k)$. By
Lemma~\ref{ext:sommerfeldscalar},
\[
 \Delta B_\ell\le \mathsf{k}\sum_k b_\ell(r_k)^{3/2}
                     \le\mathsf{k}B_\ell^{3/2}.
\]
On a positive set of $v-B_\ell$, both $v$ and $B_\ell$ are positive, and
$\Delta(v-B_\ell)>0$. The sum of barriers diverges at every finite
boundary point of $D_\ell$ and tends to zero at infinity, whereas
$v\to-\eta-A\pi^2/\ell^2<0$. The maximum principle gives
$v\le B_\ell$.

Newton's theorem for the radial unit-mass kernel gives
$|\cdot|^{-1}*g_\ell^2\le|\cdot|^{-1}$, with equality outside its
support. Thus $t\le t*g_\ell^2$ on $D_\ell$, and consequently
\begin{equation}\label{ext:sumbarrier}
 t(x)\le\frac{A\pi^2}{\ell^2}
 +\frac{M_{\rm S}^2}{\mathsf{k}^2}
   \sum_kr_k^{-4}\left[1-(\ell/r_k)^\theta\right]^{-4}.
\end{equation}
Only the smearing kernel is radial; no symmetry of the electronic density
has been used.

At a fixed evaluation point choose $\ell=\beta/H(x)$. This is smaller
than every $r_k$. Put $q_k=(r_kH(x))^{-1}$, so $q_k\ge0$ and
$\sum_kq_k=1$. Then
\[
 \sum_kr_k^{-4}[1-(\ell/r_k)^\theta]^{-4}
 =H^4\sum_k\frac{q_k^4}{[1-(\beta q_k)^\theta]^4}
 \le\frac{H^4}{(1-\beta^\theta)^4},
\]
using $q_k\le1$ and $\sum_kq_k^4\le1$.
The localization term in \eqref{ext:sumbarrier} is
$A\pi^2H^2/\beta^2$. This proves \eqref{ext:U}--\eqref{ext:ab}.
The localization error is incurred once, rather than once per nucleus.
\end{proof}

\subsection{Density comparison on harmonic level sets}

The next proposition follows the density-comparison strategy of
Solovej~\cite{Solovej}. Its differential calculation uses the exact
relativistic functions $F$ and $T$.

\begin{proposition}\label{ext:density}
If $\gamma\le1/1233$ and $9/20\le\beta\le1/2$, then
\begin{equation}\label{ext:densitybound}
 T(p(x))\le\kappa U_\beta(H(x)),\qquad \kappa=\frac{103}{100}.
\end{equation}
This estimate has no total-charge or separation hypothesis.
\end{proposition}
\begin{proof}
Abbreviate $a=a_\beta$, $b=b_\beta$, and $U=aH^4+bH^2$. Define
\[
 p_\kappa=\sqrt{\kappa U(\kappa U+2)},\qquad
 P_\kappa=F(p_\kappa),\qquad z=\frac{aH^2}{b},
\]
and put
\[
 m(z)=\frac{12z+2}{1+z},\qquad
 n(z)=\frac{4z+2}{1+z}.
\]
At $p=p_\kappa$, write
$s=\sqrt{1+p^2}$, $\mathcal H=f(p)^2/p$, and $D=pf'(p)/f(p)$.
Direct differentiation, using $\Delta H=0$, gives the exact identity
\begin{equation}\label{ext:Gidentity}
 \frac{\Delta P_\kappa}{j(p_\kappa)U}
       =\frac{\tau_H}{b}\mathcal R(p,z),
\end{equation}
where
\begin{equation}\label{ext:G}
 \mathcal R(p,z)=\frac{\mathcal H}{1+z}
 \left\{\frac{s}{s+1}m(z)
       +\frac{s^2D-1}{(s+1)^2}n(z)^2\right\}.
\end{equation}
For clarity, for any scalar function $J$, the chain rule gives
$\Delta J(H)=J''(H)|\nabla H|^2$. For its one-centre counterpart
$J(1/r)$ the radial Laplacian is $r^{-4}J''(1/r)$.
Thus the full molecular Laplacian is the corresponding one-centre
expression multiplied by $\tau_H$. Equivalently, the two coefficients
obtained by differentiating $F(T^{-1}(\kappa U))$, before division by
$U$, are
$\mathcal Hs/(s+1)$ multiplying $\Delta U/U$ and
$\mathcal H(s^2D-1)/(s+1)^2$ multiplying $|\nabla U|^2/U^2$.

We next prove $\mathcal R<4/3$ along this barrier. Since $\theta<4/5$ and
$\beta\ge9/20$,
$\beta^\theta\ge(9/20)^{4/5}>13/25$; indeed
\[
 (9/20)^4-(13/25)^5=\frac{7464617}{2500000000}>0.
\]
Using $M_{\rm S}>12$ and $9<\pi^2<10$ gives
\begin{equation}\label{ext:abBounds}
 a\alpha^2>\frac{3515625}{2048}>1600,\qquad
 b/\lambda<\frac{3440}{81}<43,\qquad
 \frac{b^2}{a}<\frac{1849}{1600}\gamma^2.
\end{equation}

If $z\ge36$, \eqref{ext:scalarbounds} and $m\le12$, $n\le4$ imply
\[
 \mathcal R\le\frac{17}{8}\frac{12+(7/10)16}{1+z}
       \le\frac{493}{370}<\frac43.
\]
Here $s/(s+1)\le1$ and $(s^2D-1)/(s+1)^2\le D$.

If $0\le z\le36$, then
\[
 w:=T(p_\kappa)=\kappa U
       =\kappa\frac{b^2}{a}z(1+z)
       <\frac{7046539}{6756840000}<\frac1{900}.
\]
Consequently $q=p_\kappa^2=w(w+2)<1801/810000<1/400$. On this range,
\begin{equation}\label{ext:smallscalar}
 \mathcal H\le1+2q,\qquad D\le\frac12+4q,\qquad
 \frac{s}{s+1}\le\frac12+\frac q8,\qquad
 \frac{s^2D-1}{(s+1)^2}\le-\frac18+\frac65q.
\end{equation}
For the derivative bound, use $h=f^2\ge p/s$ and
\[
 h'(p)=\frac{1+2q}{(1+q)^{3/2}}
       +\frac{4p\asinh p}{(1+q)^2}\le1+6q,
 \qquad
 D\le\tfrac12(1+q/2)(1+6q)\le\tfrac12+4q.
\]
The next bound uses
$s/(s+1)=1/2+q/[2(s+1)^2]$. For the last one, writing
$B=(s^2D-1)/(s+1)^2$,
\[
 B+\frac18
 =\frac{8(1+q)D-8+(s+1)^2}{8(s+1)^2}
 \le\frac{38q+32q^2}{32}
 =\frac{19}{16}q+q^2\le\frac65q.
\]
Substitution into \eqref{ext:G} gives
\[
 \mathcal R\le(1+2q)\left\{g_0(z)+\frac{207}{10}q\right\},
 \qquad g_0(z)=\frac{4z^2+5z+1/2}{(1+z)^3}.
\]
The maximum of $g_0$ occurs at $z=(\sqrt{15}-1)/4$ and is
$(40\sqrt{15}-144)/9<5/4$. Therefore
\[
 \mathcal R\le\frac{201}{200}\left(\frac54+\frac{207}{4000}\right)
       =\frac{1046607}{800000}<\frac43.
\]
This completes the scalar estimate.

At this point it is essential to retain the sign of $\mathcal R$ correctly.
There is no need to assert that $\mathcal R$ is nonnegative: by
\eqref{ext:Hidentities},
\begin{equation}\label{ext:tauG}
 \tau_H\mathcal R\le\max\{\mathcal R,0\}\le\frac43.
\end{equation}
Thus \eqref{ext:Gidentity} remains an upper bound even where the second
derivative of the scalar composite barrier is negative. Since
$\beta\le1/2$,
\[
 \frac{\lambda}{b}\le\frac{25}{86\pi^2}
       <\frac3{100}=\kappa-1.
\]
It follows that
\[
 \frac{3\lambda}{4}\Delta P_\kappa
       \le(\kappa-1)Uj(p_\kappa).
\]
On a positive set of $P-P_\kappa$, one has $p>p_\kappa$,
$T(p)>\kappa U$, and $t\le U$. The monotonicity of $j$ and
\eqref{ext:system} yield
\[
 \frac{3\lambda}{4}\Delta P
       =j(p)(T(p)-t)
       >j(p_\kappa)(\kappa-1)U
       \ge\frac{3\lambda}{4}\Delta P_\kappa.
\]
Near every nucleus $P$ is bounded and $P_\kappa\to\infty$.
At infinity both tend to zero. A positive maximum of $P-P_\kappa$ is
therefore impossible, proving \eqref{ext:densitybound}.
\end{proof}

\subsection{The exterior subharmonic function}

Define
\begin{equation}\label{ext:Bdefinition}
 I(p)=\int_0^p sf(s)^2\,\mathrm ds,\qquad g(p)^2=2I(p),\qquad
 B_\eta=\sqrt{\frac{\lambda}{\alpha\pi}g(p)^2+
                    (\varphi-\eta/\alpha)^2}-\frac\eta\alpha.
\end{equation}

\begin{lemma}\label{ext:subharmonic}
The function $B_\eta$ is subharmonic wherever $0<p\le1/2$ away from
the nuclei. Furthermore $B_\eta\ge-\varphi$, and $B_\eta\to0$ uniformly at infinity.
\end{lemma}
\begin{proof}
Put $B=B_\eta+\eta/\alpha$. The nonnegative Hessian term in the
Laplacian of a Euclidean norm, combined with \eqref{ext:system}, gives
\begin{equation}\label{ext:Bcalculation}
 B\Delta B_\eta\ge
 \frac{\lambda}{\alpha\pi}gg'
       \left(\frac{g''}{g'}-\frac{f'}f\right)|\nabla p|^2
       +\frac{4p^3T(p)}{3\alpha\pi}.
\end{equation}
The potential terms cancel exactly because $gg'=pf^2$.
To check the remaining gradient sign, set
$e(p)=ph'(p)/h(p)$, where $h=f^2$, and use
$v=p/\sqrt{1+p^2}$ and $w_{\rm h}(v)=v\atanh v$.
Differentiation gives
\[
 e=\frac{1+v^2+4(1-v^2)w_{\rm h}}{1+2w_{\rm h}},\qquad
 \frac{\mathrm de}{\mathrm dv}
 =\frac{2\{v(1-2w_{\rm h}-8w_{\rm h}^2)+(1-3v^2)w_{\rm h}'\}}
        {(1+2w_{\rm h})^2}.
\]
For $p\le1/2$, one has $v^2\le1/5$,
$0\le w_{\rm h}\le v^2/(1-v^2)=p^2\le1/4$, and $w_{\rm h}'>0$.
Thus $e$ is increasing. As $(p^2h)'=ph(2+e)$, integration gives
\[
 I(p)\ge\frac{p^2h(p)}{2+e(p)}.
\]
For $\mathcal K(p)=p^2h(p)/I(p)$ this implies
$p\mathcal K'/\mathcal K=2+e-\mathcal K\ge0$. Since
$g''/g'-f'/f=\tfrac12(\log\mathcal K)'$, the first term on the right of
\eqref{ext:Bcalculation} is nonnegative.

Finally,
$B_\eta\ge|\varphi-\eta/\alpha|-\eta/\alpha\ge-\varphi$,
and the stated limit follows from $p,\varphi\to0$ and $\eta\ge0$.
In particular $B_\eta$ need not be nonnegative.
\end{proof}

\section{Quartic coercivity with nuclear repulsion}
\label{coer:section}

We now use the total energy \eqref{eq:energy}, including $U_{\rm nn}$.
We prove the following bound, retaining a positive multiple of $\int p^4$
with constants
independent of the number and positions of nuclei. This argument modifies
the localization strategy of \cite{Siedentop}; the negative Hardy correction
is explicitly estimated.

\begin{proposition}[Quartic coercivity]\label{prop:coercivity}
For every $\lambda>0$, $0<\alpha\le1/137$, and $0<\alpha Z_k\le1$,
every admissible density satisfies
\begin{equation}\label{coer:general_charges}
 \E[p;\mathbf R,\mathbf Z]
 \ge\frac1{16\pi^2}\int p^4-N-\alpha C_\lambda Z_{\rm tot},
 \qquad C_\lambda=\frac8\pi\sinh\!\left(\frac{16}{3\lambda}\right).
\end{equation}
Consequently, for $E_+=\max\{\E[p;\mathbf R,\mathbf Z],0\}$ and $M>0$,
\begin{equation}\label{coer:high_density_mass}
 N_{\{p\ge M\}}:=\frac1{3\pi^2}\int_{\{p\ge M\}}p^3
 \le\frac{16}{3M}\bigl(N+\alpha C_\lambda Z_{\rm tot}+E_+\bigr).
\end{equation}
\end{proposition}

We give the proof in four steps. None of these steps requires electronic
minimality or the sign condition \eqref{eq:sublevel}.

\subsection{The electrostatic inequality, with its normalization}

Suppose first that all nuclear charges equal $Z>0$ and $K\ge2$.
Write $\Gamma_k$ for the Voronoi cell of $R_k$ and
\[
 r_k(x)=|x-R_k|,\qquad
 D_k=\frac12\min_{l\ne k}|R_k-R_l|,\qquad B_k=B(R_k,D_k).
\]
The balls $B_k$ are disjoint and are contained in their respective cells.
The electrostatic inequality of Lieb--Yau, in the form also used by
Lieb--Loss--Siedentop~\cite[Eq.~(3.11)]{LLS}, gives
\begin{equation}\label{coer:electrostatic}
 \D[\rho]-Z\sum_k\int\frac{\rho(x)}{r_k(x)}\,dx
 +Z^2\sum_{k<l}\frac1{|R_k-R_l|}
 \ge
 -Z\sum_k\int_{\Gamma_k}\frac{\rho(x)}{r_k(x)}\,dx
 +\frac{Z^2}{8}\sum_k\frac1{D_k}.
\end{equation}
For clarity, here is a derivation of precisely this normalization.
Set
\[
 V=\sum_k\frac Z{r_k},\qquad
 W=\max_k\frac Z{r_k},\qquad \Phi=V-W,
 \qquad U=Z^2\sum_{k<l}|R_k-R_l|^{-1}.
\]
The nuclear singularities of $\Phi$ cancel.  The distribution
$\sigma=-\Delta\Phi/(4\pi)$ is a nonnegative measure supported on the
Voronoi faces, and $\Phi=\sigma*|\cdot|^{-1}$.  On a face shared by
$\Gamma_k$ and $\Gamma_l$, with $d_{kl}=|R_k-R_l|$ and $r_k=r_l=r$,
\[
 d\sigma=\frac{Zd_{kl}}{4\pi r^3}\,dS.
\]
The measure $\sigma$ has finite mass: its density on each of the finitely
many faces is integrable. Moreover, $\Phi$ is bounded, and
$\Phi(x)=(K-1)Z/|x|+O(|x|^{-2})$ at infinity. The flux of $-\nabla\Phi$
through large spheres therefore gives $\sigma(\R^3)=(K-1)Z$.
Consequently,
\[
 \D[\sigma]=\frac12\int\Phi\,d\sigma
 \le\frac12\|\Phi\|_\infty(K-1)Z<\infty.
\]
Thus the Coulomb square is well defined, and its nonnegativity gives
\[
 \D[\rho]-\int\Phi\rho\ge-\D[\sigma].
\]
The two distributional identities
\[
 -\Delta\Phi=4\pi\sigma,
 \qquad -\Delta W=4\pi\sum_kZ\delta_{R_k}-4\pi\sigma
\]
give, by integration by parts,
\[
 4\pi\int W\,d\sigma
 =\int\nabla W\cdot\nabla\Phi
 =8\pi\bigl(U-\D[\sigma]\bigr).
\]
These identities can first be justified with small balls about the nuclei
removed and a large exterior cutoff, then by passage to the limit.
It follows that
\begin{align*}
 U-\D[\sigma]
 &=\frac12\int W\,d\sigma
 =\frac{Z^2}{8\pi}\sum_{\text{faces }k,l}
       \int\frac{d_{kl}}{r_k^4}\,dS\\
 &=\frac{Z^2}{8\pi}\sum_k
       \int_{\partial\Gamma_k}
       \frac{n_k\cdot(x-R_k)}{r_k^4}\,dS.
\end{align*}
In the last equality, the two contributions of a shared face sum to
$d_{kl}/r_k^4$.
The divergence theorem on $\Gamma_k\setminus B_k$ yields
\[
 \int_{\partial\Gamma_k}
       \frac{n_k\cdot(x-R_k)}{r_k^4}\,dS
 =\frac{4\pi}{D_k}
   -\int_{\Gamma_k\setminus B_k}\frac{dx}{r_k^4}.
\]
The cell lies in the half-space bounded by the bisector to a nearest
neighbour.  At radius $r\ge D_k$, the solid angle of that half-space is
$2\pi(1+D_k/r)$.  Therefore
\begin{equation}\label{coer:voronoi_integral}
 \int_{\Gamma_k\setminus B_k}\frac{dx}{r_k^4}
 \le\int_{D_k}^{\infty}2\pi(r^{-2}+D_kr^{-3})\,dr
 =\frac{3\pi}{D_k}.
\end{equation}
Thus $U-\D[\sigma]\ge(Z^2/8)\sum_kD_k^{-1}$, proving
\eqref{coer:electrostatic}.  No restriction on $Z$, $K$, or the separations
is used in this electrostatic lemma.

\subsection{A signed local Hardy inequality}

For $0<r_k<D_k$ define
\[
 A_k(x)=\left(\frac1{2r_k}-\frac{r_k}{2D_k^2}\right)
             \frac{x-R_k}{r_k},
\]
and extend $A_k$ by zero outside $B_k$.  Its normal component vanishes at
$\partial B_k$.  Expanding the nonnegative square
$\int|\nabla v_s+(\sum_kA_k)v_s|^2$ gives
\begin{equation}\label{coer:signed_hardy}
 \int|\nabla v_s|^2
 \ge\sum_k\int_{B_k}
   \left[\frac1{4r_k^2}
         -\frac1{D_k^2}Y\!\left(\frac{r_k}{D_k}\right)\right]v_s^2,
 \qquad Y(t)=1+\frac{t^2}{4}.
\end{equation}
Indeed,
\[
 \operatorname{div}A_k-|A_k|^2
 =\frac1{4r_k^2}-\frac1{D_k^2}-\frac{r_k^2}{4D_k^4}.
\]
The inequality holds for $v_s\in H^1(\mathbb R^3)$ by approximation.
The negative part in \eqref{coer:signed_hardy} is essential and is retained.

\subsection{Truncation of the density and explicit coercivity}

Put $\zeta=\alpha Z$ and choose
\begin{equation}\label{coer:constants}
 s=\sinh\!\left(\frac{16\zeta^2}{3\lambda}\right),\qquad
 v_s=(p-s)_+,\qquad
 a_{\rm W}=\frac{2\zeta^2}{\pi^2},\qquad
 C=\frac{8s\zeta^3}{\pi},\qquad
 A_{\rm loc}=\frac{27308}{945\pi}.
\end{equation}
For all $p\ge0$,
\[
 f(p)^2=\frac p{\sqrt{1+p^2}}
       +\frac{2p^2}{1+p^2}\asinh p
 \ge\asinh p.
\]
For example, $p\sqrt{1+p^2}\ge p\ge\asinh p$ bounds the
first term from below by $\asinh p/(1+p^2)$.
Consequently
\[
 T_W\ge\frac{3\lambda}{8\pi^2}\asinh(s)
                     \int|\nabla v_s|^2
      =a_{\rm W}\int|\nabla v_s|^2.
\]
For $s>0$ the truncation belongs to $H^1$: its gradient is controlled by
this estimate, and $v_s^2\le s^{-1}p^3$ on $\{p>s\}$ gives its $L^2$ bound.
The elementary kinetic bound
$\tau(p)\ge2p^4-\frac83p^3$ gives
\[
 T_{\mathrm{TF}}\ge\frac1{4\pi^2}\int p^4-N.
\]
Retain $\frac1{16\pi^2}\int p^4$.  In each ball, allocate the remaining
quartic term as $p^4/(8\pi^2)+p^4/(16\pi^2)$; outside the balls it is
$3p^4/(16\pi^2)$.  Combining \eqref{coer:electrostatic} and
\eqref{coer:signed_hardy} leaves three estimates.

\smallskip\noindent\textit{The singular term in a ball.}\quad
On $p\ge3s$ one has $v_s\ge2p/3$, and hence
\[
 \frac{a_{\rm W}v_s^2}{4r_k^2}+\frac{p^4}{8\pi^2}
    -\frac{\zeta p^3}{3\pi^2r_k}
 \ge\frac1{8\pi^2}
       \left(p^2-\frac{4\zeta p}{3r_k}\right)^2\ge0.
\]
On $p<3s$, discard the nonnegative Hardy term.  Extend the pointwise
negative lower bound to all of $\mathbb R^3$.  The minimizing value is
$p_{\min}(r)=\min\{3s,2\zeta/r\}$, with transition radius
$r_*=2\zeta/(3s)$.  Direct integration gives
\begin{equation}\label{coer:low_density}
 \int_{B_k\cap\{p<3s\}}
   \left(\frac{p^4}{8\pi^2}
         -\frac{\zeta p^3}{3\pi^2r_k}\right)dx
 \ge-\frac{8s\zeta^3}{\pi}=-C.
\end{equation}
The radial integrals inside and outside $r_*$ each equal
$-4s\zeta^3/\pi$.

\smallskip\noindent\textit{The negative Hardy correction.}\quad
Use $v_s^2\le p^2$ and pointwise minimization to obtain
\[
 \frac{p^4}{16\pi^2}
 -\frac{a_{\rm W}}{D_k^2}Y(r_k/D_k)v_s^2
 \ge-\frac{16\zeta^4}{\pi^2D_k^4}Y(r_k/D_k)^2.
\]
Since
\[
 \int_0^1r^2(1+r^2/4)^2\,dr=\frac{743}{1680},
\]
the integrated correction is at least
\begin{equation}\label{coer:hardy_error}
 -\frac{64\zeta^4}{\pi D_k}\frac{743}{1680}
 =-\frac{2972\zeta^4}{105\pi D_k}.
\end{equation}

\smallskip\noindent\textit{The exterior of a ball in its Voronoi cell.}\quad
Pointwise minimization gives
\[
 \frac{3p^4}{16\pi^2}-\frac{\zeta p^3}{3\pi^2r_k}
 \ge-\frac{16\zeta^4}{81\pi^2r_k^4}.
\]
By \eqref{coer:voronoi_integral}, its integral is at least
\begin{equation}\label{coer:exterior_error}
 -\frac{16\zeta^4}{27\pi D_k}.
\end{equation}
Combining \eqref{coer:low_density}--\eqref{coer:exterior_error} proves
\begin{equation}\label{coer:equal_charge}
 \E[p;\mathbf R,(Z,\ldots,Z)]\ge\frac1{16\pi^2}\int p^4-N-CK
 +\left(\frac{\alpha Z^2}{8}-A_{\rm loc}\zeta^4\right)\sum_k\frac1{D_k}.
\end{equation}
For $K=1$ the same estimate holds with the distance term omitted, using
the ordinary global Hardy inequality and the same high/low truncation.
For $K=0$, $\mathcal E\ge(16\pi^2)^{-1}\int p^4-N$ is immediate.

\subsection{Individual and unequal nuclear charges}

Suppose $0<\alpha\le1/137$ and $0<\alpha Z\le1$. The distance coefficient
in \eqref{coer:equal_charge} is nonnegative, because $A_{\rm loc}<10$ and
\[
 \frac{\alpha Z^2}{8}-A_{\rm loc}\zeta^4
 =\zeta^2\left(\frac1{8\alpha}-A_{\rm loc}\zeta^2\right)
 \ge\zeta^2\left(\frac{137}{8}-A_{\rm loc}\right)>0.
\]
The constant in \eqref{coer:constants} satisfies
$C\le C_\lambda=(8/\pi)\sinh(16/(3\lambda))$.
After discarding the distance term, the equal-charge bound is therefore
\[
 \E\ge\frac1{16\pi^2}\int p^4-N-C_\lambda K.
\]

For unequal charges, put $Z_*=1/\alpha$ and take independent random variables
$\xi_k$ equal to $Z_*$ with probability $\alpha Z_k$, and zero otherwise.
For each fixed density and geometry the energy is separately affine in the
nuclear charges. Independence gives
\[
 \mathbb E\,\E[p;\mathbf R,\boldsymbol\xi]
 =\E[p;\mathbf R,\mathbf Z].
\]
For each realization, let $K_{\boldsymbol\xi}=\#\{k:\xi_k=Z_*\}$.
The active nuclei all have charge $Z_*$. Applying the preceding
equal-charge bound with $Z=Z_*$ and $K=K_{\boldsymbol\xi}$ gives
\[
 \E[p;\mathbf R,\boldsymbol\xi]
 \ge\frac1{16\pi^2}\int p^4-N-C_\lambda K_{\boldsymbol\xi}.
\]
The cases of one or no active nuclei were covered above. Taking
expectations in this inequality and using
\[
 \mathbb E K_{\boldsymbol\xi}
 =\sum_{k=1}^K\mathbb P(\xi_k=Z_*)
 =\alpha\sum_{k=1}^K Z_k=\alpha Z_{\rm tot}
\]
proves
\eqref{coer:general_charges}. Finally, $p^3\le p^4/M$ on $\{p\ge M\}$
gives \eqref{coer:high_density_mass} and completes the proof of
Proposition~\ref{prop:coercivity}.

\section{A nonlinear comparison with an integrable source}\label{sec:nonlinear}

We continue to assume only $0<\gamma=\alpha\lambda\le\gamma_0=1/1233$
until the coercivity estimate is invoked in Section~\ref{sec:charge}.
The comparison function below agrees with a multiple of $P^{4/3}$ at small
density and with its tangent thereafter. Convexity preserves the sign of
the gradient term in the Euler equation. The scalar comparison holds up
to a large explicit momentum; any failure above it is represented by a
Newton potential.

\subsection{The scalar comparison and its high-density cutoff}

\begin{lemma}[A convex comparison with a tangent continuation]
\label{newG:scalar}
Let
\[
 0<\gamma=\alpha\lambda\le\gamma_0:=\frac1{1233},\qquad
 L=\gamma_0/\gamma\ge1,\qquad p_0=4\gamma,\qquad P_0=F(p_0),
\]
and put
\[
 \beta_0=\frac34,\qquad
 c=\frac{\beta_0}{2}\left(\frac32\right)^{4/3},\qquad
 m=\frac{4c}{3}P_0^{1/3},\qquad b=\frac c3P_0^{4/3}.
\]
Define on $[0,\infty)$
\begin{equation}\label{newG:definition}
 G(P)=
 \begin{cases}
  cP^{4/3},&0\le P\le P_0,\\
  mP-b,&P\ge P_0.
 \end{cases}
\end{equation}
Then $G$ is nonnegative, convex, and continuously differentiable, with
$G(0)=G'(0)=0$, $0<G'(P)\le m$ for $P>0$, and $G(P)\le mP$.
The function
\begin{equation}\label{newG:defect}
 D_G(p)=G(F(p))+\frac{\gamma p f(p)}{\pi G'(F(p))}-T(p)
 \quad(p>0)
\end{equation}
extends continuously by $D_G(0)=0$. With
\[
 d=\frac{361}{1000}\gamma^2,\qquad p_{\rm cut} = e^{180L},
\]
one has
\begin{equation}\label{newG:scalarbound}
 D_G(p)\le d\quad(0\le p\le p_{\rm cut}),
 \qquad
 D_G(p)<0\quad(p_0\le p\le p_{\rm cut}).
\end{equation}
Moreover,
\begin{equation}\label{newG:slope}
 \frac32\sqrt\gamma\le m\le\frac{15001}{10000}\sqrt\gamma,
 \qquad b\ge2\gamma^2.
\end{equation}
\end{lemma}

\begin{proof}
The two expressions in \eqref{newG:definition} have the same value and
derivative at $P_0$. Their derivative increases from zero to $m$ and
then remains constant. This proves the asserted regularity and
convexity. The inequality $G(P)\le mP$ follows directly on each branch.

We first record the elementary small-$p$ bounds
\begin{equation}\label{newG:smallf}
 \sqrt p\le f(p)\le\sqrt p(1+p^2),\qquad
 \frac23p^{3/2}\le F(p)
 \le\frac23p^{3/2}\left(1+\frac37p^2\right)
 \quad(0\le p\le1/100).
\end{equation}
The upper bounds follow from $f(p)^2\le p+2p^3$. To prove the lower
bound, use $\asinh p\ge p/\sqrt{1+p^2}$, which gives
\[
 f(p)^2\ge p\frac{1+3p^2}{(1+p^2)^{3/2}}\ge p
 \qquad(0\le p\le1).
\]
Indeed, for $x=p^2\in[0,1]$,
$(1+3x)^2-(1+x)^3=x(3+6x-x^2)\ge0$.
Integration proves the lower bound on $F$. We shall also use
\begin{equation}\label{newG:smallT}
 T(p)\ge\frac12p^2-\frac18p^4.
\end{equation}
This follows from
$T(p)=p^2/(\sqrt{1+p^2}+1)\ge p^2/(2+p^2/2)$ and
$(1+v)^{-1}\ge1-v$.

Since $p_0\le4/1233<1/100$, \eqref{newG:smallf} applies at $p_0$.
Substitution into $m$ and $b$ gives
\[
 \frac32\le\frac m{\sqrt\gamma}
 \le\frac32\left(1+\frac37p_0^2\right)^{1/3},
 \qquad b\ge\frac18p_0^2=2\gamma^2.
\]
Set
\[
 \Delta_0=\frac{16}{1233^2}=\frac{16}{1520289},\qquad
 M_0=\frac{15001}{10000},\qquad \pi_-=\frac{333}{106}<\pi.
\]
The upper slope bound follows from the exact positive gap
\[
 M_0^3-\left(\frac32\right)^3
            \left(1+\frac37\Delta_0\right)
 =\frac{86689437366383}{131383000000000000}>0.
\]

\smallskip\noindent\textit{The power branch: $0\le p\le p_0$.}\quad
For $0\le x\le3$ one has
$(1+3x/7)^{4/3}\le1+x$; this follows, for example, by differentiating
$\log(1+x)-(4/3)\log(1+3x/7)$.
Using \eqref{newG:smallf} therefore gives
\[
 \frac38p^2\le G(F(p))
 \le\frac38(1+p_0^2)p^2,
 \qquad G'(F(p))\ge\frac34\sqrt p.
\]
Together with \eqref{newG:smallT},
\[
 D_G(p)
 \le-\left(\frac18-\frac12p_0^2\right)p^2
      +\frac{\gamma(1+p_0^2)}{\pi\beta_0}p
 \le-A_0p^2+B_0\gamma p,
\]
where
\[
 A_0=\frac18-\frac12\Delta_0>0,
 \qquad B_0=\frac{1+\Delta_0}{\pi_-\beta_0}.
\]
The maximum of the last quadratic on $p\ge0$ is
$B_0^2\gamma^2/(4A_0)$. The exact bound is
\[
 \frac{B_0^2}{4A_0}
 =\frac{33241694034468992}{92262451740651801}
 <\frac{361}{1000},
\]
with gap
\[
 \frac{361}{1000}-\frac{B_0^2}{4A_0}
 =\frac{65051043906308161}{92262451740651801000}>0.
\]
The same estimates give the stated continuous extension at $p=0$:
$G(F(p))=O(p^2)$, $G'(F(p))\sim\beta_0\sqrt p$, and the quotient
term in \eqref{newG:defect} is $O(\gamma p)$.

\smallskip\noindent\textit{The initial tangent branch: $p_0\le p\le1/100$.}\quad
Here
\[
 D_G(p)=mF(p)-b+\frac{\gamma p f(p)}{\pi m}-T(p).
\]
The preceding estimates imply
\[
 D_G(p)\le C_0\sqrt\gamma\,p^{3/2}
     -\frac{39999}{80000}p^2-2\gamma^2,
\]
where
\[
 C_0=M_0\left(\frac23+\frac2{70000}\right)
       +\frac{1+1/10000}{\pi_-\,(3/2)}
 <\frac{49}{40}.
\]
For reproducibility, $49/40-C_0=632219143/49950000000>0$.
Writing $u=p/\gamma\ge4$, we obtain
\[
 \frac{D_G(p)}{\gamma^2}
 \le h(u):=\frac{49}{40}u^{3/2}
       -\frac{39999}{80000}u^2-2.
\]
One has $h(4)=-999/5000<0$ and
\[
 h'(u)=\sqrt u\left(\frac{147}{80}
                -\frac{39999}{40000}\sqrt u\right)<0
 \quad(u\ge4).
\]
This proves strict negativity throughout this interval.

\smallskip\noindent\textit{The remaining tangent branch.}\quad
The global scalar inequality $F(p)\le pf(p)$ gives
\begin{equation}\label{newG:tailestimate}
 D_G(p)\le\sqrt\gamma\,C_m\,pf(p)-T(p),\qquad
 C_m:=\frac m{\sqrt\gamma}+\frac{\sqrt\gamma}{\pi m}.
\end{equation}
By \eqref{newG:slope},
\[
 C_m\le M_0+\frac1{\pi_-\,(3/2)}<\frac{12}{7},
 \qquad
 \frac{12}{7}-M_0-\frac1{\pi_-\,(3/2)}
 =\frac{138007}{69930000}>0.
\]
For $1/100\le p\le1/10$, use
$T(p)\ge399p^2/800$ and
$f(p)\le(101/100)\sqrt p$ to obtain
\[
 \frac{T(p)}{pf(p)}\ge\frac{399}{8080}
 >\frac{12}{7}\sqrt{\gamma_0}.
\]
The last comparison follows by squaring, with exact gap
\[
 \left(\frac{399}{8080}\right)^2
  -\left(\frac{12}{7}\right)^2\frac1{1233}
 =\frac{24133913}{438267603200}>0.
\]
On $[1/10,1]$ and $[1,5]$ the required lower bounds are,
respectively,
\[
 \frac{T(p)}{pf(p)}\ge\frac2{25},\qquad
 \frac{T(p)}{pf(p)}\ge\frac2{21}.
\]
For clarity these follow from $f(p)^2/p\le17/8$, using
$\sqrt{17/8}<3/2$, $\sqrt{1/10}>3/10$,
$\sqrt2+1<5/2$, and $\sqrt{26}+1<7$ in
\[
 \frac{T(p)}{pf(p)}
 \ge\frac{\sqrt p}{\sqrt{17/8}(\sqrt{1+p^2}+1)}.
\]
Both constants exceed $(12/7)\sqrt{\gamma_0}$. Thus
\eqref{newG:tailestimate} is strictly negative up to $p=5$.

For $5\le p\le100$, the estimates
\[
 f(p)^2\le1+2\asinh p<13,
 \qquad \frac{T(p)}p>1-\frac1p\ge\frac45
\]
give strict negativity. The sufficient squared comparison is
\[
 \left(\frac45\right)^2
 -\left(\frac{12}{7}\right)^2\frac{13}{1233}
 =\frac{102208}{167825}>0.
\]
Here $\asinh 100<\log300<6$, and
$\log300<6$ follows from $e>8/3$ and $(8/3)^6>300$.

Finally, for $100\le p\le e^{180L}$,
\[
 f(p)^2\le1+2\asinh p
 <1+2\log(3p)<360L+4\le364L.
\]
Since $\gamma L=\gamma_0$, one obtains
\[
 C_m\sqrt\gamma f(p)
 <\frac{12}{7}\sqrt{\frac{364}{1233}}
 <\frac{99}{100}<\frac{T(p)}p.
\]
The exact middle squared gap is
\[
 \left(\frac{99}{100}\right)^2
 -\left(\frac{12}{7}\right)^2\frac{364}{1233}
 =\frac{1079159}{9590000}>0.
\]
This proves \eqref{newG:scalarbound} on the full asserted interval.
\end{proof}

\subsection{The global elliptic comparison and its Newton source}

\begin{lemma}[Nonlinear comparison with a Newton source]\label{lem:newtoncomparison}
Put
\begin{gather}
 q=(d-\eta)_+,\qquad
 c_G(p)=\frac{4G'(F(p))}{3\lambda}\frac{p^2}{f(p)},
 \qquad c_G(0)=0,\label{newG:coefficient}\\
 S(x)=c_G(p(x))(D_G(p(x))-d)_+,\qquad
 \Psi(x)=\frac1{4\pi}\int\frac{S(y)}{|x-y|}\,dy.
 \label{newG:source}
\end{gather}
Then $S$ is nonnegative, bounded, and compactly supported, with
$S=0$ on $\{p\le p_{\rm cut}\}$, and
\begin{equation}\label{newG:globalcomparison}
 \boxed{\quad G(P)\le\alpha\varphi+q+\Psi.\quad}
\end{equation}
No upper restriction on $\eta\ge0$ is required.
\end{lemma}

\begin{proof}
Near zero, $c_G(p)=O(p^2)$, so its stated extension is continuous.
The qualitative boundedness and decay of $p$, together with
\eqref{newG:scalarbound}, give all claims about $S$. Its Newton potential
satisfies $\Psi\ge0$, $-\Delta\Psi=S$, and $\Psi\in C_0(\mathbb R^3)$.

The convex chain rule gives
\[
 \Delta G(P)=G'(P)\Delta P+G''(P)|\nabla P|^2
 \ge G'(P)\Delta P
\]
in distributions. To justify it directly, on each compact set $P$ has
a positive lower bound, and $G'$ is Lipschitz on the corresponding
range of $P$. One can use $P\in W^{2,2}_{\rm loc}$ and the ordinary
Sobolev chain rule, or approximate $G$ by smooth convex functions.
There is no surface measure at $P=P_0$, because the two derivatives
in \eqref{newG:definition} agree there. The value assigned to $G''$ at that
level is immaterial. The possible singularity of $G''$ at $P=0$ does
not affect a compact-set argument for a positive minimizer.

Set $w=G(P)-\alpha\varphi-q$. The Euler and Poisson equations yield
\begin{align}
 (-\Delta+c_G(p))w
 ={}&c_G(p)(D_G(p)-\eta-q)
      -G''(P)|\nabla P|^2
      -4\pi\alpha\sum_kZ_k\delta_{R_k}\\
 \le{}&S.
 \label{newG:wequation}
\end{align}
Here the exact cancellation is
\[
 c_G(p)\frac{\gamma p f(p)}{\pi G'(F(p))}
 =\frac{4\alpha}{3\pi}p^3=4\pi\alpha\rho,
\]
and $\eta+q=\max\{\eta,d\}\ge d$.
The term involving $G''$ has the favorable sign. No derivative of
$c_G$ is taken: it is only a zeroth-order coefficient.
Since
\[
 (-\Delta+c_G(p))\Psi=S+c_G(p)\Psi\ge S,
\]
we have $(-\Delta+c_G(p))(w-\Psi)\le0$.
At a positively charged nucleus, $w-\Psi\to-\infty$ because $G(P)$
and the electronic Coulomb potential are bounded. At infinity,
$w-\Psi\to-q\le0$. The maximum principle on punctured bounded
domains, followed by exhaustion, gives \eqref{newG:globalcomparison}.
The coefficient $c_G\ge0$ is bounded; compact support of this
coefficient is neither asserted nor required.
\end{proof}

\begin{lemma}[Mass of the nonlinear defect]\label{lem:sourcemass}
With $N_{\rm high}=\int_{\{p>p_{\rm cut}\}}\rho$, one has
\begin{equation}\label{newG:mass}
 \frac1{4\pi\alpha}\int S\le9N_{\rm high}.
\end{equation}
\end{lemma}

\begin{proof}
On the support of $S$ we have $p>e^{180L}>4\gamma$, and therefore
$G'(F(p))=m$. As $T,d\ge0$ and $G(F(p))\le mF(p)\le mpf(p)$,
\begin{align*}
 S&\le\frac{4mp^2}{3\lambda f(p)}
       \left(m+\frac\gamma{\pi m}\right)pf(p)\,
       \mathbf1_{\{p>p_{\rm cut}\}}\\
  &=4\pi\alpha\left(1+\frac{\pi m^2}{\gamma}\right)
       \rho\,\mathbf1_{\{p>p_{\rm cut}\}}\\
  &\le36\pi\alpha\rho\,\mathbf1_{\{p>p_{\rm cut}\}}.
\end{align*}
The last inequality follows from
$1+(22/7)(15001/10000)^2<9$. Integration proves the claim.
\end{proof}

\section{The excess-charge estimate}\label{sec:charge}

\subsection{Boundary comparison, with the chemical potential retained}

We use $G,d,q,S,\Psi$, and $p_{\rm cut}$ from Section~\ref{sec:nonlinear}.
Set
\begin{equation}\label{gb:parameters}
 x_*=\frac{66}{5},\quad \beta=\frac{47}{100},\quad
 r=\frac{x_*}{\alpha\sqrt\lambda},\quad
 b_*=\frac{2239}{5000},\quad
 \Omega_r=\{x:H(x)<1/r\}.
\end{equation}

\begin{proposition}[Exterior boundary estimate]\label{prop:boundary}
For $0<\gamma\le\gamma_0$, throughout $\overline\Omega_r$ one has
\begin{equation}\label{gb:boundary}
 \alpha B_\eta\le b_*\gamma^2+\Psi,
 \qquad \alpha B_\eta=\sqrt{(\gamma/\pi)g(p)^2+t^2}-\eta.
\end{equation}
The estimate holds for every $\eta\ge0$.
\end{proposition}
\begin{proof}
Write $d=\delta\gamma^2$, where $\delta=361/1000$.
\smallskip\noindent\textit{Certified exterior density and potential bounds.}\quad
Use Propositions~\ref{ext:potentialbarrier} and~\ref{ext:density} with
$\kappa=103/100$ and $M_{\rm S}=692881/57600$.  Put
\[
 \overline V=\frac{76607}{250000},\qquad
 z_* =\frac{79451}{100000},\qquad
 c_3=\frac23\left(1+\frac1{10^6}\right).
\]
The rational choices $\pi_-=333/106$, $\pi_+=355/113$, and
$\overline\sigma=55703/100000$ satisfy
$\beta^{31}<\overline\sigma^{40}$ and
\[
 \pi_+^2\left[
 \frac{9M_{\rm S}^2}{128x_*^4(1-\overline\sigma)^4}
 +\frac{43}{50\beta^2x_*^2}\right]<\overline V,
\]
\[
 \kappa\overline V(2+\kappa\gamma_0^2\overline V)<z_*^2.
\]
For $x\in\overline\Omega_r$, one has $H(x)\le1/r$ and hence
\[
 t(x)\le U_\beta(H(x))\le U_\beta(1/r)<\gamma^2\overline V,
 \qquad T(p(x))\le\kappa U_\beta(H(x))<\kappa\gamma^2\overline V.
\]
Using $p^2=T(p)(T(p)+2)$ and $\gamma\le\gamma_0$, we obtain
\[
 p(x)^2\le\gamma^2\kappa\overline V
             \bigl(2+\kappa\gamma_0^2\overline V\bigr)
          <\gamma^2z_*^2.
\]
Thus, throughout the closed exterior,
\begin{equation}\label{gb:exteriorcaps}
 t\le\gamma^2\overline V,\qquad
 p<\gamma z_*<\gamma<4\gamma.
\end{equation}
In particular, $G$ is evaluated on its small-density branch there.
The usual estimate $f(p)^2\le p+2p^3$ gives
\[
 g(p)^2\le\frac23p^3\left(1+\frac65p^2\right)\le c_3p^3,
\]
because $(6/5)\gamma_0^2<10^{-6}$.

The lower bound in \eqref{newG:smallf} also gives
$G(P)\ge3p^2/8$ on this exterior.

\smallskip\noindent\textit{The positive-$t$ part.}\quad
For $t\ge0$, the bounds above give
\[
 (\gamma/\pi)g(p)^2+t^2
 <\gamma^4\left[(c_3/\pi_-)z_*^3+\overline V^2\right]
 <b_*^2\gamma^4.
\]
The last inequality is a strict rational inequality; the exact expressions
for this and the following numerical margins are recorded in
Appendix~\ref{app:arithmetic}. Hence $\alpha B_\eta<b_*\gamma^2-\eta$, which implies
\eqref{gb:boundary}.

\smallskip\noindent\textit{The negative-$t$ part, uniformly in the multiplier.}\quad
At a fixed point let
\[
 a=\eta+q+\Psi=\max\{\eta,d\}+\Psi\ge d.
\]
The comparison \eqref{newG:globalcomparison} and the lower bound for $G$ yield
\[
 0\le\frac38p^2\le t+a.
\]
If $t<0$, this implies $-a\le t<0$ and
\[
 (\gamma/\pi)g(p)^2+t^2
 \le\frac{\gamma c_3}{\pi}
          \left[\frac83(t+a)\right]^{3/2}+t^2.
\]
The right-hand side is a convex function of $t$ on $[-a,0]$.
Its maximum is attained at one of the endpoints.  Therefore
\[
 \sqrt{(\gamma/\pi)g(p)^2+t^2}\le M(a),\qquad
 M(a)=\max\{a,C\sqrt\gamma\,a^{3/4}\},
\]
where $C=(c_3/\pi)^{1/2}(8/3)^{3/4}$.

The following two rational checks are sufficient:
\[
 b_*>\delta,\qquad
 b_*^4>(c_3/\pi_-)^2(8/3)^3\delta^3,
\]
\[
 1-\frac{3b_*}{4\delta}=\frac{503}{7220}>0.
\]
The first line gives $M(d)<b_*\gamma^2$.  For $a\ge d$, the derivative
of the second branch is bounded by
\[
 \frac34C\sqrt\gamma\,d^{-1/4}
 <\frac{3b_*}{4\delta}<1.
\]
The first branch has derivative one. Both branches are therefore
increasing and $1$-Lipschitz on $[d,\infty)$, and so is their maximum.
It follows pointwise, without differentiating $\Psi$, that
\begin{align*}
 \alpha B_\eta
 &\le M(a)-\eta\le M(d)+a-d-\eta\\
 &<b_*\gamma^2+\Psi-\min\{\eta,d\}\le b_*\gamma^2+\Psi.
\end{align*}
This proves \eqref{gb:boundary} for every $\eta\ge0$.
\end{proof}

\subsection{Harmonic comparison and extraction of the charge}

\begin{proposition}[Charge estimate before absorption]\label{prop:rawcharge}
For $0<\gamma\le\gamma_0$ one has
\begin{equation}\label{newG:rawQ}
 Q\le \frac{147774}{25000}\lambda^{3/2}K
       +\frac1{4\pi\alpha}\int S
 \le\frac{147774}{25000}\lambda^{3/2}K+9N_{\rm high},
\end{equation}
where $N_{\rm high}=\int_{\{p>p_{\rm cut}\}}\rho$.
\end{proposition}
\begin{proof}
The density cap \eqref{gb:exteriorcaps} gives $p<\gamma<1/2$ on
$\overline\Omega_r$. Lemma~\ref{ext:subharmonic} shows that
$B_\eta\ge-\varphi$, $B_\eta\to0$ at infinity, and $\Delta B_\eta\ge0$
in this exterior. Since $-\Delta\Psi=S\ge0$, the function
$B_\eta-\Psi/\alpha$ is subharmonic there. By
Proposition~\ref{prop:boundary}, its boundary values are at most
$b_*\gamma^2/\alpha$. The harmonic comparison therefore gives
\begin{equation}\label{eq:harmoniccomparison}
 -\varphi\le B_\eta\le\frac{b_*\gamma^2}{\alpha}\,rH+\frac\Psi\alpha
 \qquad\hbox{on }\Omega_r.
\end{equation}
For a direct exhaustion argument, first take a regular level
$h<1/r$ of $H$. The maximum principle on
$\{H<h\}$, intersected with a large ball, compares
$B_\eta-\Psi/\alpha$ with $(b_*\gamma^2/\alpha)H/h$.
All functions vanish at infinity, so the outer boundary error tends to
zero. Let regular values $h\uparrow1/r$ to obtain
\eqref{eq:harmoniccomparison}. This also handles a nonregular boundary
of $\Omega_r$.

Let $\langle u\rangle_R=(4\pi)^{-1}\int_{\mathbb S^2}u(R\omega)\,d\omega$.
For $R$ enclosing every nucleus, Newton's theorem gives
\begin{align}
 R\langle H\rangle_R&=K,\label{eq:NewtonH}\\
 -R\langle\varphi\rangle_R
 &=\int\min\{1,R/|y|\}\rho(y)\,dy-Z_{\rm tot}
 \longrightarrow Q,\label{eq:Newtonphi}\\
 R\langle\Psi\rangle_R
 &=\frac1{4\pi}\int\min\{1,R/|y|\}S(y)\,dy
 \longrightarrow\frac1{4\pi}\int S.\label{eq:NewtonPsi}
\end{align}
The limits use only the integrability of $\rho$ and $S$.
For sufficiently large $R$, the whole sphere lies in $\Omega_r$.
Average \eqref{eq:harmoniccomparison}, multiply by $R$, and let
$R\to\infty$. Since
$b_*\gamma^2r/\alpha=x_*b_*\lambda^{3/2}$ and
$x_*b_*=147774/25000$, the first inequality in \eqref{newG:rawQ} follows.
The second is Lemma~\ref{lem:sourcemass}.
\end{proof}

\subsection{Absorption and proof of Theorem~\ref{thm:main}}

\begin{proof}[Proof of Theorem~\ref{thm:main}]
If $p=0$, then $Q=-Z_{\rm tot}<0$ and the assertion is immediate.
For a nonzero minimizer, Lemma~\ref{lem:reduction} reduces the proof to
\eqref{eq:normalizeddomain}. We first establish the result in this domain.
Since $\E\le0$, Proposition~\ref{prop:coercivity} gives
\[
 N_{\rm high}\le\frac{16}{3p_{\rm cut}}
      \bigl(N+\alpha C_\lambda Z_{\rm tot}\bigr),
 \qquad C_\lambda=\frac8\pi\sinh\!\left(\frac{16}{3\lambda}\right).
\]
With $\varepsilon=48e^{-180L}$, $N=Z_{\rm tot}+Q$, and
$Z_{\rm tot}\le K/\alpha$, \eqref{newG:rawQ} therefore yields
\begin{equation}\label{newG:absorb}
 (1-\varepsilon)Q
 \le\left[
   \frac{147774}{25000}\lambda^{3/2}
   +\varepsilon\left(\frac1\alpha+C_\lambda\right)
 \right]K.
\end{equation}
In particular, $\varepsilon<1$. Put $B_{\rm out}=5913/1000$.
The parameter hypotheses imply
\[
 \lambda^{-1}\le9L,\qquad C_\lambda<2e^{48L},\qquad
 \frac1{\alpha\lambda^{3/2}}\le3699L^{3/2},\qquad
 \frac{C_\lambda}{\lambda^{3/2}}
 <54L^{3/2}e^{48L}.
\]
Since $L^{3/2}e^{-aL}$ is decreasing for $L\ge1$ when $a\ge3/2$,
\begin{align*}
 \varepsilon\left[
 B_{\rm out}+\frac1{\alpha\lambda^{3/2}}
                  +\frac{C_\lambda}{\lambda^{3/2}}
 \right]
 &<48(B_{\rm out}+3699)e^{-180}+2592e^{-132}\\
 &<2^{-162}+2^{-120}<2^{-119}<2^{-9}\\
 &<\frac{51}{25000}
   =B_{\rm out}-\frac{147774}{25000}.
\end{align*}
The powers of two use $e>2$,
$48(B_{\rm out}+3699)<2^{18}$, and $2592<2^{12}$.
Combining this strict margin with \eqref{newG:absorb} proves
\begin{equation}\label{newG:finalbound}
 \boxed{\displaystyle
 Q<\frac{5913}{1000}\lambda^{3/2}K.}
\end{equation}
At $\lambda=1/9$, this is exactly
\begin{equation}\label{newG:physicalbound}
 \boxed{\displaystyle Q<\frac{219}{1000}K.}
\end{equation}

For the original invariant parameter domain, apply the change of variables
in Lemma~\ref{lem:reduction}. Electronic minimality, finiteness of the mass,
and the energy sign are preserved; multiplication of the transformed
bound by $s^3$ proves \eqref{eq:main}.
\end{proof}

\begin{remark}[A positive-energy correction]\label{rem:positiveenergy}
The same proof under \eqref{eq:normalizeddomain} gives, without the sign
restriction on the total energy,
\[
 Q<\frac{5913}{1000}\lambda^{3/2}K
      +\frac{\varepsilon}{1-\varepsilon}E_+,
 \qquad
 \varepsilon=48\exp\!\left(-\frac{180}{1233\alpha\lambda}\right).
\]
Indeed, \eqref{coer:high_density_mass} adds $\varepsilon E_+$ to
\eqref{newG:absorb}. By covariance, the same formula holds under
\eqref{eq:invariantdomain}: $E_+$ and $Q$ acquire the same scaling factor,
while $\varepsilon$ is invariant. This correction can be used for
electronically minimizing configurations whose positive total energies
tend to zero. No claim about existence or convergence of such a sequence
is needed for the estimate.
\end{remark}

\section{Nuclear configurations and the factor \texorpdfstring{$K$}{K}}\label{sec:nuclear}

\subsection{A mass-preserving dilation}

The following variation keeps the nuclear charges and the parameters
$\alpha,\lambda$ fixed.  It is therefore distinct from a covariance
identity that changes the parameters of the functional.  Let $p$ be an
admissible finite-mass density variable and set, for $s>0$,
\begin{equation}\label{vir:variation}
 p_s(x)=s\,p(sx),\qquad R_{k,s}=s^{-1}R_k.
\end{equation}
Then $\rho_s(x)=s^3\rho(sx)$, and consequently
$\int\rho_s=\int\rho$.  In particular, the Lagrange multiplier for the
electron number contributes nothing to this variation.

Write $P=F(p)$ and
\[
 \vartheta(p)=\frac{pf'(p)}{f(p)},\qquad
 c_\lambda=\frac{3\lambda}{8\pi^2},\qquad
 c_0=\frac1{8\pi^2},
\]
where $\vartheta(0)$ is defined by its continuous limit.  Denote the full Coulomb
part of the energy by
\[
 \mathcal C=-\alpha\int V\rho+\alpha\D[\rho]
       +\alpha\sum_{k<l}\frac{Z_kZ_l}{|R_k-R_l|}.
\]
Every term in $\mathcal C$, including nuclear repulsion, is multiplied
by $s$ under \eqref{vir:variation}.  The exact energy along this path is
\begin{equation}\label{vir:energy_path}
 \mathcal E[p_s;\mathbf R_s,\mathbf Z]
 =c_\lambda s\int
       \frac{f(sp)^2}{f(p)^2}|\nabla P|^2
   +c_0s^{-3}\int\tau(sp)+s\mathcal C.
\end{equation}
At $p=0$ the ratio in the first integral is defined by continuity and
equals $s$.  This form of the identity requires no unweighted
$H^1$ regularity of $p$.

Here are sufficient bounds to justify differentiating
\eqref{vir:energy_path}.  The previously established scalar estimate
$0<\vartheta(p)\le7/10$ gives, uniformly for $s$ in a compact subinterval of
$(0,\infty)$, a bound on $f(sp)^2/f(p)^2$ and on the derivative of
$s f(sp)^2/f(p)^2$.  Thus the gradient integrands are dominated by a
constant times $|\nabla P|^2$.  The Sobolev chain rule applies because
$u\mapsto F(sF^{-1}(u))$ has bounded derivative for each such range of
$s$.

For the local kinetic term one has
\begin{equation}\label{vir:tau_elasticity}
 4\tau(p)\le p\tau'(p)\le5\tau(p).
\end{equation}
Indeed, for $p>0$,
\[
 \frac{pT'(p)}{T(p)}=1+\frac1{\sqrt{1+p^2}}\in[1,2].
\]
It follows, for $0\le u\le1$, that
$u^2T(p)\le T(up)\le uT(p)$.  Substitution in
\[
 \tau(p)=8p^3\int_0^1u^2T(up)\,du
\]
gives $p\tau'(p)/5\le\tau(p)\le p\tau'(p)/4$, proving
\eqref{vir:tau_elasticity}.  This estimate also bounds
$\tau(sp)$ and the derivative of $s^{-3}\tau(sp)$ by a constant times
$\tau(p)$ when $s$ stays in a compact subinterval of $(0,\infty)$.
Dominated convergence therefore justifies the differentiation.

\begin{proposition}[Virial identity and the energy sign]\label{prop:virial}
If an admissible nonzero pair $(p,\mathbf R)$ is stationary along the
dilation \eqref{vir:variation}, then
\begin{equation}\label{vir:negative_energy}
 \E[p;\mathbf R,\mathbf Z]
 =-\frac{3\lambda}{4\pi^2}\int\vartheta(p)|\nabla P|^2
  -\frac1{8\pi^2}\int\bigl(p\tau'(p)-4\tau(p)\bigr)<0.
\end{equation}
\end{proposition}
\begin{proof}
The preceding bounds justify differentiation at $s=1$ and give
\begin{equation}\label{vir:identity}
 0=c_\lambda\int(1+2\vartheta(p))|\nabla P|^2
    +c_0\int\bigl(p\tau'(p)-3\tau(p)\bigr)+\mathcal C.
\end{equation}
Subtracting this identity from the energy proves the equality in
\eqref{vir:negative_energy}. Both integrands are nonnegative. In fact,
for $h(p)=p\tau'(p)-4\tau(p)$ one has $h(0)=0$ and
\[
 h'(p)=8p^2\left(1-\frac1{\sqrt{1+p^2}}\right)>0\qquad(p>0).
\]
Thus the local integral is strictly positive for a nonzero density.
\end{proof}

\subsection{Local minima of the Born--Oppenheimer energy}

Fix one of the mass constraints admitted in Section~\ref{sec:model} and
define the electronic ground-state energy, including nuclear repulsion, by
\[
 E_{\rm BO}(\mathbf R)=\inf_p\E[p;\mathbf R,\mathbf Z].
\]

\begin{corollary}[Locally minimizing nuclear configurations]\label{cor:BO}
Suppose that $\mathbf R$ is a local minimum of $E_{\rm BO}$, that the
electronic infimum at $\mathbf R$ is attained by a nonzero finite-mass
density $p$, and that \eqref{eq:invariantdomain} holds. Then
\[
 \E[p;\mathbf R,\mathbf Z]<0,\qquad
 N-Z_{\rm tot}<\frac{5913}{1000}\lambda^{3/2}K.
\]
In particular, the conclusion holds for any attained joint ground state
with nonzero electronic density.
\end{corollary}
\begin{proof}
The dilation preserves the electron number, and hence the chosen mass
constraint. For $s$ near one,
\[
 \E[p_s;\mathbf R/s,\mathbf Z]
 \ge E_{\rm BO}(\mathbf R/s)
 \ge E_{\rm BO}(\mathbf R)
 =\E[p;\mathbf R,\mathbf Z].
\]
Its derivative at one vanishes. Proposition~\ref{prop:virial} and
Theorem~\ref{thm:main} give the assertions.
\end{proof}

This argument requires neither differentiability of $E_{\rm BO}$ nor
uniqueness of its electronic minimizer. The electronic minimizing
hypothesis remains essential to the charge theorem. The corollary is
conditional on attainment and does not require an existence theorem for
nuclear minimizing configurations.

If a dissociation limit consists of attained minimizing fragments, the
theorem can be applied separately to each fragment satisfying its energy
hypothesis. Summing gives the same coefficient times the sum of their
numbers of nuclei. A fragment cannot be omitted merely because its total
energy is positive: the present theorem then supplies only the correction
in Remark~\ref{rem:positiveenergy}. This observation makes no assertion
about the existence of a particular fragmentation.

\subsection{The harmonic geometry of the factor \texorpdfstring{$K$}{K}}\label{vir:capacity}

The harmonic comparison uses the function $H$, harmonic away from the nuclei,
and the associated level sets
\[
 H(x)=\sum_{k=1}^K\frac1{|x-R_k|},\qquad
 E_r=\{x:rH(x)\ge1\},\qquad
 \Omega_r=\mathbb R^3\setminus E_r.
\]
In the Newtonian capacity convention $\operatorname{cap}(B(0,a))=a$,
the function $\min\{1,rH\}$ is the capacitary potential of $E_r$:
it belongs to $D^{1,2}(\mathbb R^3)$, is superharmonic, equals one on
$E_r$, is harmonic on its complement, and tends to zero at infinity.
Since
\[
 rH(x)=\frac{Kr}{|x|}+O(|x|^{-2})\qquad(|x|\to\infty),
\]
one has exactly
\begin{equation}\label{vir:capacity_identity}
 \operatorname{cap}(E_r)=Kr.
\end{equation}
This remains true when components of $E_r$ merge.  A homothety rescales
capacity and $r$ by the same factor and hence does not improve $K$ in
this comparison.  A smaller geometric coefficient would require a
different comparison set together with adequate estimates of the
density and potential on its boundary; it does not follow from scaling
the present equipotential construction.

\appendix
\section{The localization coefficient}\label{app:localization}

We prove the scalar estimate \eqref{ext:Aconstant} used in
Lemma~\ref{ext:formlemma}. The argument is algebraic and holds for every
$p>0$.

Put
$u_{\rm loc}=\asinh p$, $c=\cosh u_{\rm loc}$,
$s=\sinh u_{\rm loc}$, and
\[
 U_0=(c+2u_{\rm loc}s)^2,\qquad
 D_0=c(3c^2-s^2)+4u_{\rm loc}s(c^2+s^2)>0.
\]
Direct differentiation gives
\[
 \frac34\frac{f(p)^2}{p(2-pf'/f)}=\frac32\frac{U_0}{D_0}.
\]
It therefore suffices to prove $J(u_{\rm loc}):=43D_0-75U_0>0$. Hyperbolic
identities yield
\begin{align*}
 J(u_{\rm loc})={}&\frac{43}{2}\cosh(3u_{\rm loc})
       +\frac{215}{2}\cosh u_{\rm loc}
       -\frac{75}{2}-\frac{75}{2}\cosh(2u_{\rm loc})\\
 &+86u_{\rm loc}\bigl(\sinh(3u_{\rm loc})-\sinh u_{\rm loc}\bigr)
       -150u_{\rm loc}\sinh(2u_{\rm loc})\\
 &-150u_{\rm loc}^2\cosh(2u_{\rm loc})+150u_{\rm loc}^2.
\end{align*}
For $n\ge2$, its coefficients
$a_n=(2n)![u_{\rm loc}^{2n}]J(u_{\rm loc})$ are
\[
 a_n=A_n-B_n-C_n,\quad
 A_n=\frac{129+344n}{6}9^n,\quad
 B_n=\frac{300n^2+150n+75}{2}4^n,\quad
 C_n=172n-\frac{215}{2}.
\]
One has $a_3=37453>0$; for $n\ge3$,
$A_{n+1}>9A_n$, $B_{n+1}\le9B_n$, and $C_{n+1}\le9C_n$.
The second inequality reduces to $20n^2-22n-19\ge0$.
Consequently all $a_n$ with $n\ge3$ are positive, and
$J(u_{\rm loc})\ge P_5(u_{\rm loc}^2)$, where
\begin{align*}
 P_5(y)={}&54-\frac{105}{2}y-\frac{1807}{24}y^2
 +\frac{37453}{720}y^3
 +\frac{314779}{13440}y^4
 +\frac{13933781}{3628800}y^5.
\end{align*}
For $y=1+v\ge1$,
\begin{align*}
 3628800P_5(1+v)={}&19912031+238970785v+942350750v^2\\
 &+668062250v^3+154659235v^4+13933781v^5>0.
\end{align*}
For $0\le y\le1$, expand $P_5$ in the degree-five Bernstein basis on
each of the following intervals. The minimum Bernstein coefficients are
\[
 \begin{array}{c|c}
 [a,b]&\min_{0\le j\le5}b_j\\\hline
 {[0,1/2]}&1975594121/116121600\\
 {[1/2,3/4]}&116537703/45875200\\
 {[3/4,1]}&78069/409600
 \end{array}
\]
and are strictly positive. These entries are obtained by substituting
$y=a+(b-a)v$ and using
$b_j=\sum_{k=0}^j q_k\binom jk/\binom5k$ when
$P_5(a+(b-a)v)=\sum_{k=0}^5q_kv^k$.
This proves $P_5>0$ and hence \eqref{ext:Aconstant}.

\section{Finite arithmetic checks}\label{app:arithmetic}

The formulae of this appendix have been checked by exact rational
arithmetic in standard-library Python.

The bounds on $\pi$ used throughout are
\[
 \pi_-:=\frac{333}{106}<\pi<\frac{355}{113}=: \pi_+.
\]
They may be certified using Machin's identity
$\pi=16\arctan(1/5)-4\arctan(1/239)$ and the alternating series
\[
 A_n(x)=\sum_{\nu=0}^{n-1}\frac{(-1)^\nu x^{2\nu+1}}{2\nu+1}.
\]
For $0<x<1$, $A_{2n}(x)<\arctan x<A_{2n+1}(x)$; substitution with
$n=5$ in each term already gives rational bounds inside $(\pi_-,\pi_+)$.

For the boundary estimate let
\[
 \delta=\frac{361}{1000},\quad b_*=\frac{2239}{5000},\quad
 \overline V=\frac{76607}{250000},\quad z_*=\frac{79451}{100000},\quad
 c_3=\frac23(1+10^{-6}),\quad
 \overline\sigma=\frac{55703}{100000}.
\]
With $x_*=66/5$, $\beta=47/100$, $\kappa=103/100$, and
$M_{\rm S}=692881/57600$, define the following rational margins:
\begin{align*}
 \Delta_V&=\overline V-\pi_+^2
  \left[\frac{9M_{\rm S}^2}{128x_*^4(1-\overline\sigma)^4}
       +\frac{43}{50\beta^2x_*^2}\right],\\
 \Delta_p&=z_*^2-\kappa\overline V(2+\kappa\gamma_0^2\overline V),\\
 \Delta_+&=b_*^2-\left[(c_3/\pi_-)z_*^3+\overline V^2\right],\\
 \Delta_-&=b_*^4-(c_3/\pi_-)^2(8/3)^3\delta^3.
\end{align*}
The exact positivity checks can equivalently be bounded below by
\[
 \Delta_V>9\cdot10^{-7},\quad
 \Delta_p>4\cdot10^{-6},\quad
 \Delta_+>19\cdot10^{-5},\quad
 \Delta_->3\cdot10^{-5}.
\]
The remaining boundary checks are
$\beta^{31}<\overline\sigma^{40}$,
$(6/5)\gamma_0^2<10^{-6}$, $b_*>\delta$, and
$1-3b_*/(4\delta)=503/7220>0$.
The localization Bernstein coefficients are displayed in
Appendix~\ref{app:localization}; the scalar cutoff proof records its
interval estimates directly. Finally, the absorption uses the exact gap
\[
 \frac{5913}{1000}-\frac{147774}{25000}=\frac{51}{25000}>
 2^{-9},\qquad
 \frac1{27}\frac{5913}{1000}=\frac{219}{1000}.
\]
These formulae specify the finite arithmetic independently of the software.

\end{document}